\documentclass{elsart}

\usepackage{amssymb}
\usepackage{amsmath}
\usepackage{bm}
\usepackage{latexsym}
\usepackage{graphicx}
\usepackage{times}
\DeclareMathAlphabet{\mathpzc}{OT1}{pzc}{m}{it}
\DeclareMathAlphabet{\mathcalligra}{T1}{calligra}{m}{n}

\begin{document}

\begin{frontmatter}
\title{A simplified nonlinear memory function\\ for the dynamics of glass-forming materials \\based on time-convolutionless mode-coupling theory}

\author{Michio Tokuyama}
\address{Institute of Multidisciplinary Research for Advanced Materials, Tohoku University, Sendai 980-8577, Japan}

\date{\today}

\begin{abstract}
A simplified nonlinear memory function is proposed in the ideal time-convolutionless mode-coupling theory equation to study the dynamics of glass-forming liquids. The numerical solutions are then compared with the simulation results performed on fragile liquids and strong liquids. They are shown to recover the simulation results in a supercooled state well within error, except at a $\beta$-relaxation stage because of the ideal equation. A temperature dependence of the nonlinearity $\mu$ in the memory function then suggests that the supercooled state must be clearly separated into two substates, a weakly supercooled state in which $\mu$ increases rapidly as $T$ decreases and a deeply supercooled state in which $\mu$ becomes constant up to the glass transition as $T$ decreases. On the other hand, it is shown that in a glass state $\mu$ increases rapidly as $T$ decreases, while it is constant in a liquid state. Thus, it is emphasized that the new model for the simplified memory function is much more reasonable than the conventional one proposed earlier by the present author not only qualitatively but also quantitatively.
\end{abstract}

\begin{keyword}
Glass transition; Simplified nonlinear memory function; Supercooled liquids; Time-convolutionless mode-coupling theory; Universality

\maketitle
\end{keyword}
\end{frontmatter}

\section{Introduction}
Understanding of the glass transition from first principles is one of the important works left in condensed matters physics. A considerable attention has been drawn to study the dynamics of glass-forming materials near the
glass transition experimentally and theoretically \cite{mct84L,mct84B,mct91,f93,yip95,ca,ds,ngai,d03,bk,got,to,bb11,toku14,mm16}. 

In order to study the glass transition from a statistical-mechanical point of view, we have recently proposed the time-convolutionless mode-coupling theory (TMCT) \cite{toku14} and then derived the second-order differential equation for the cumulant function $K(q,t)(=-\ln[f(q,t)])$ \cite{toku15}, where $f(q,t)$ is the scaled collective-intermediate scattering function. We call this an ideal TMCT equation. This equation contains the nonlinear memory function $\Delta\varphi(q,t)$, which has the same form as that obtained in the mode-coupling theory (MCT) \cite{mct84L,mct84B,mct91}. The memory function is written in terms of the static structure factor $S(k)$ and is integrated over the wavevector $\bm{k}$. Once the analytic form of $S(k)$ is known, the ideal TMCT equation is easily solved numerically. Here it is well-known that depending on temperature $T$, there exist three characteristic states in the glass-forming liquids, a liquid state [L], a supercooled state [S], and a glass state [G]. It is also well-known that depending on a time scale, there are four characteristic stages in the dynamics of supercooled liquids, a first stage for a short time, a $\beta$-relaxation stage for an intermediate time where $f(q,t)$ is described by the von Schweilder decay \cite{vsc}, an $\alpha$-relaxation stage for a later time where $f(q,t)$ is well-known to obey the Kohlrausch-Williams-Watts (KWW) function (or a stretched exponential decay) \cite{kw,ww}, and a diffusion stage for a long time where $f(q,t)$ obeys a diffusion equation. In order to check whether the numerical solutions of the ideal TMCT equation obey such relaxation processes or not, we have recently solved it numerically for hard spheres \cite{kim16,naru17,toku19} by using the Percus-Yevick static structure factor \cite{py}. Thus, it has been shown that the critical volume fraction $\phi_c\simeq 0.5817$ agrees with that predicted from the molecular-dynamics simulations \cite{toku03,toku031,toku05,toku071} and also that the numerical solutions describe the simulation results well, except at $\beta$ stage because of an ideal equation \cite{toku19,toku171}. Since $S(k)$ is in general not known, however, one has to obtain its numerical values from either the simulations or the experiments. This is not an easy task to do. Therefore, the main purpose of the present paper is to propose a reasonably simplified nonlinear memory function and thus to check whether the numerical solutions of the ideal TMCT equation with such a memory function can describe the simulation results performed on (F) fragile liquids and (S) strong liquids well or not. 

In the previous paper \cite{toku15}, we have first simplified the memory function $\Delta\varphi(q,t)$ around $q=q_m$ as $\Delta\varphi(q_m,t)=\Delta\varphi(q_m,0)f(q_m,t)^2$ by just employing the same approach as that used in MCT \cite{mct84B}, where $q_m$ is a first peak position of $S(q)$. Then, we have shown that the numerical solutions can roughly describe the simulation results only in [L]. In order to describe the dynamics of supercooled liquids, therefore, we have next proposed the renormalized simplified memory function \cite{toku17}
\begin{equation}
\Delta\varphi(q_m,t)=\Delta\varphi(q_m,0)f(q_m,t)^{\varepsilon}=\Delta\varphi(q_m,0)e^{-\varepsilon K(q_m,t)},\label{rsmf}
\end{equation}
where $\varepsilon$ is a nonlinear exponent to be determined. We call this a conventional model. Thus, it has been shown that $\varepsilon$ is constant to be 2.0 in [L], while it grows from 2.0 in [S] as temperature decreases. The numerical solutions have been shown to describe the simulation results performed on  two types of glass-forming liquids (F) and (S) qualitatively but not quantitatively. Especially, this model fails in describing the simulation results at $\alpha$ stage near the glass transition quantitatively. In fact, the stretched (or KWW) exponent $\beta$ obtained from the numerical solutions does not coincide with that obtained from the simulation results near the glass transition \cite{toku191}.

\begin{figure}
\begin{center}
\begin{tabular}{cc}
\begin{minipage}{.5\textwidth}
\includegraphics[width=6.7cm]{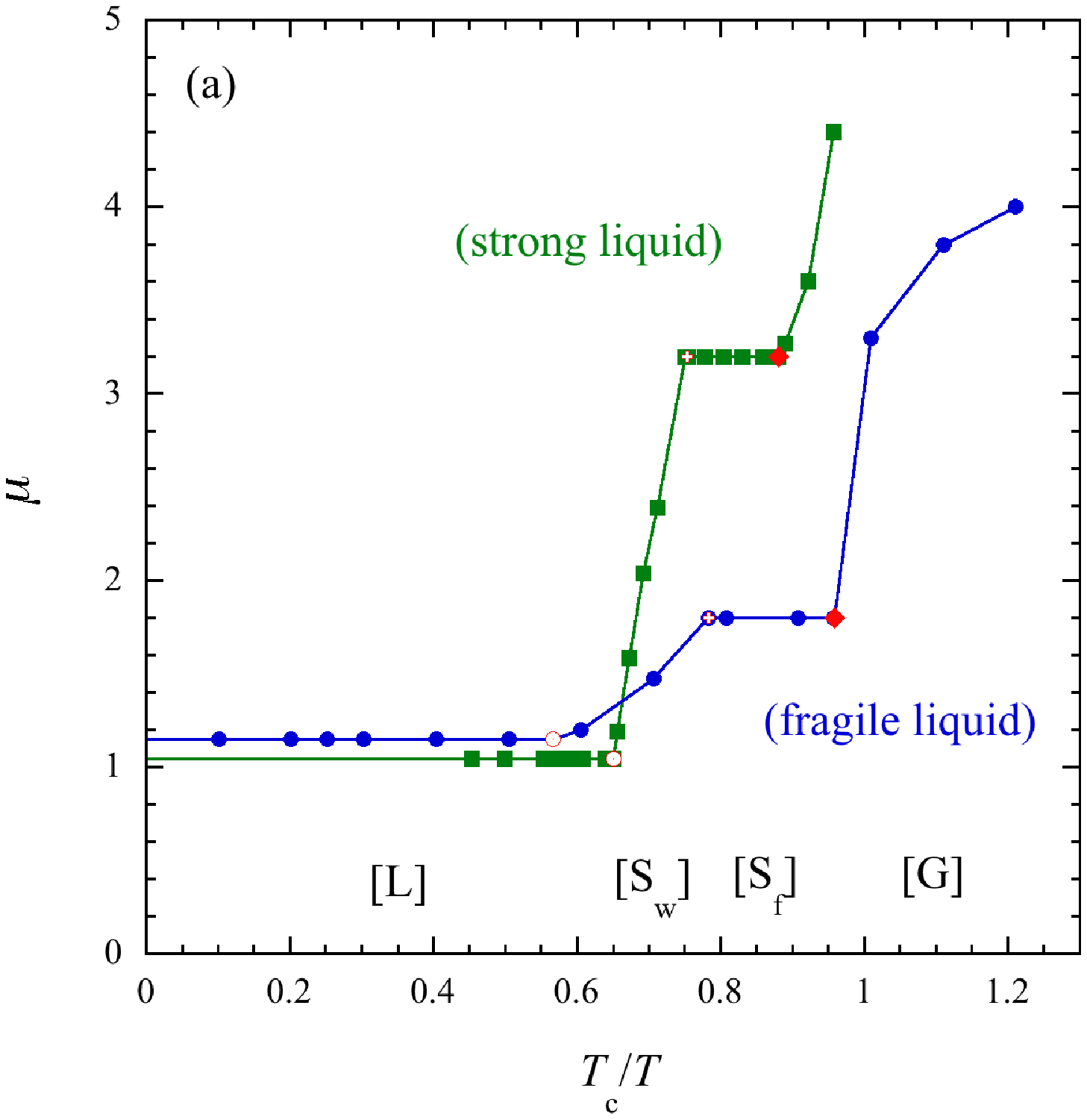}
\end{minipage}
\begin{minipage}{.5\textwidth}
\includegraphics[width=6.7cm]{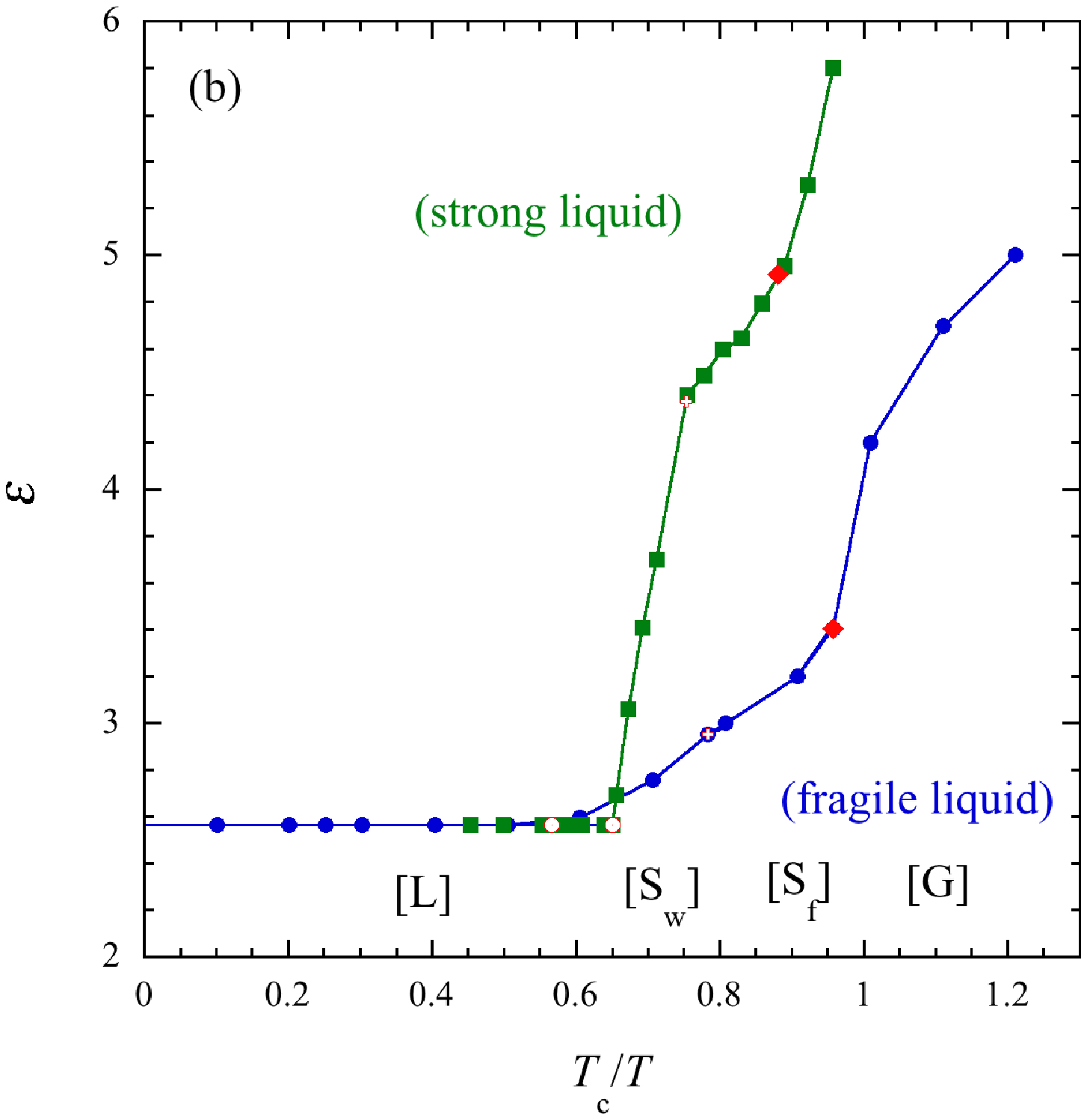}
\end{minipage}
\end{tabular}
\end{center}
\caption{(Color online) A plot of the nonlinear parameters $\mu$ and $\varepsilon$ versus scaled inverse temperature $T_c/T$ for (a) a new simplified model given by Eq. (\ref{almod}) and (b) a conventional model given by Eq. (\ref{rsmf}), where $T_c$ is a critical temperature. The symbols ($\bullet)$ indicate the fitting values for a fragile liquid and ($\Box$) for a strong liquid. The symbols ($\Diamond$) indicate the glass transition point $T_g$, ($\odot$) the supercooled point $T_s$, and ($+$) the deeply supercooled point $T_f$, where the numerical values of $T_i$ are listed in Table \ref{table-2}. The solid lines are guides for eyes.  The label [L] stands for a liquid state, [S$_w$] for a weakly supercooled state, [S$_f$] for a deeply supercooled state, and [G] for a glass state.}
\label{nexT}
\end{figure}
In order to simplify the memory function, we have assumed in the previous papers \cite{toku15,toku17} that the dominant contribution in $\Delta\varphi(q,t)$ results from the first peak $q_m$ of $S(k)$ and then simplified $S(k)$ in terms of $\delta$ function as $S(k)=1+A\delta(k-q_m)$ \cite{mct84B}, where $A$ is a positive constant to be determined. However, the disagreement between the numerical solutions and the simulation results at $\alpha$ stage suggests that the integration over the wavevector $\bm{k}$ contained in $\Delta\varphi(q,t)$ must plays an important role in finding a reasonable value of the KWW exponent $\beta$ at $\alpha$ stage. Considering such a wavevector dependence, therefore, we now propose the following simplified nonlinear memory function: 
\begin{equation}
\Delta\varphi(q_m,t)=\Delta\varphi(q_m,0)
\frac{f(q_m,t)}{[1+\mu K(q_m,t)]^{\beta/b}}, \label{almod}
\end{equation}
where $\mu$ is a nonlinear parameter to be determined and $b$ the von Schweidler exponent \cite{vsc}. We call this a new model. In Ref. \cite{toku171}, we have pointed out theoretically that the solutions of the ideal TMCT equation with the original (non-simplified) memory function deviate from the simulation results at $\beta$ stage because of the approximation employed (see next section for details). In fact, this has been checked numerically by using the PY static structure factor \cite{toku19,toku171}. Hence we note that the comparison of the numerical solutions with the simulation results must be done except at $\beta$ stage. Because of this reason, in this paper we solve the TMCT equation numerically not only for the new model given by Eq. (\ref{almod}) but also for the conventional model given by Eq. (\ref{rsmf}) under the same initial conditions obtained from the simulation results and compare both solutions together with the simulation results consistently. Thus, it turns out from the new model that the supercooled state should be further separated into two substates, a weakly supercooled state [S$_w$] and a deeply supercooled state [S$_f$]. As is shown in Fig. \ref{nexT}(a), this is because as temperature decreases, the nonlinear parameter $\mu$ becomes constant in a deeply supercooled state [S$_f$], while it grows rapidly in [S$_w$] and [G], where in [L] $\mu$ is constant to be 1.150 for (F) and 1.046 for (S). Similar behavior is seen for two types of glass-forming liquids, (F) and (S). It is also shown that the new model enables us to recover the same value of $\beta$ as that obtained for the simulation results near the glass transition \cite{toku191}. On the other hand, the conventional model shows in Fig. \ref{nexT}(b) that the nonlinear exponent $\varepsilon$ grows rapidly in [S] and [G] as temperature decreases, while $\varepsilon$ is constant to be 2.564 in [L]. 

In Section 2 we summarize the basic equations briefly, which are used in the present paper. In Section 3 we derive the universal TMCT equation, which has only one solution for different glass-forming liquids of type $i$, where $i=$(F) and (S). In Section 4 we first briefly summarize the simulation results performed on two types of glass-forming liquids, (F) and (S). Then, we solve the universal TMCT equation for the new model numerically under the initial conditions obtained from the simulation results. Thus, we show that the numerical solutions agree with the simulation results well within error, except at $\beta$ stage. We also compare them with those obtained for the conventional model under the same initial conditions consistently. We conclude in Section 5 with a summary.

\section{Basic equations}
In this section, we briefly summarize the macroscopic equations for the scaled collective-intermediate scattering function $f(q,t)$, which have been derived from first principles by using a new formulation based on TMCT \cite{toku14,toku15,toku171,toku17}. 

We consider the three-dimensional equilibrium glass-forming system, which consists of $N$ particles with mass $m$ and diameter $\sigma$ in the total volume $V$ at temperature $T$. Let $\xi$ denote the control parameter, such as a volume fraction $\phi(=\pi\rho \sigma^3/6)$ and an inverse temperature $1/T$, where $\rho(=N/V)$ is the number density. Then, the scaled collective-intermediate scattering function $f(q,t)$ is given by
\begin{equation}
f(q,t)=\langle \rho(\bm{q},t)\rho(\bm{q},0)^*\rangle/S(q) \label{cisf}
\end{equation}
with the collective-density fluctuation
$\rho(\bm{q},t)=N^{-1/2}[\sum_{j=1}^Ne^{i\bm{q}\cdot\bm{X}_j(t)}-N\delta_{\bm{q},0}]$,
where $\bm{X}_j(t)$ indicates the position vector of a $j$th particle at time $t$, the brackets the average over an equilibrium ensemble, $S(q)(=\langle|\rho(\bm{q},0)|^2\rangle)$ the static structure factor, $q=|\bm{q}|$, and $f(q,0)=1$. 

\begin{figure}
\begin{center}
\includegraphics[width=12.0cm]{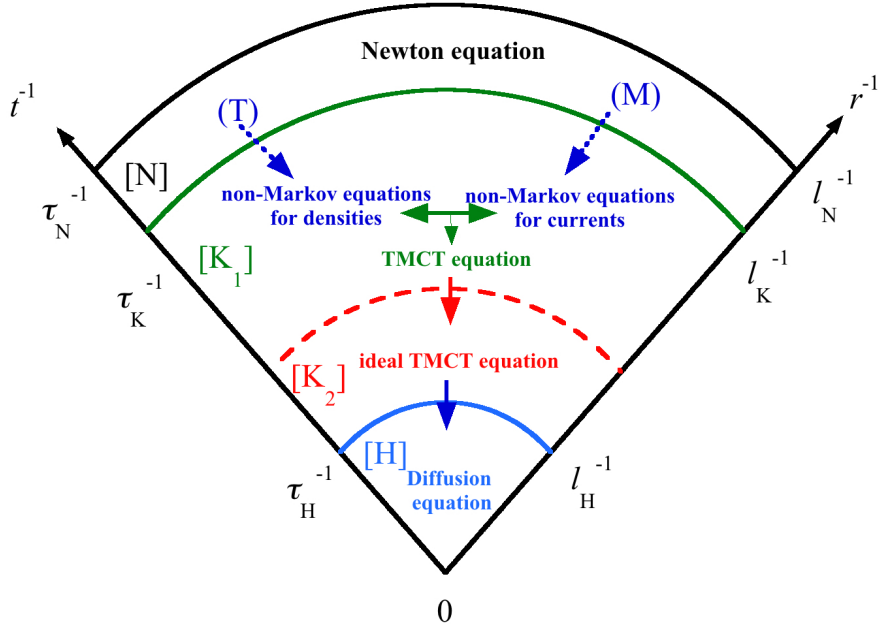}
\end{center}
\caption{(Color online) Classification of the basic equations discussed in the present paper into three stages, [N], [K], and [H], depending on a space ($r$)-time ($t$) scale, where $\ell_i$ and $\tau_i$ indicate the relevant length and time of interest, respectively.}
\label{evo}
\end{figure}
\subsection{Basic equation in each characteristic stage}
Depending on a space-time scale, there exist three characteristic stages, a microscopic stage [N], a kinetic stage [K], and a hydrodynamic stage [H] (see Fig. \ref{evo}). In a microscopic stage [N], the position $\bm{X}_j(t)$ and the momentum $\bm{P}_j(t)$ of $j$th particle at time $t$ are described by the Newton (or Heisenberg) equations. Depending on a space-time scale, the kinetic stage [K] further consists of two substages, a fast stage [K$_1$] and a slow stage [K$_2$]. In a fast stage [K$_1$], the relevant variables are given by  the number density $\rho(\bm{q},t)$ and the current density $j(\bm{q},t)(\propto d\rho(\bm{q},t)/dt)$. As discussed in the previous papers \cite{toku14,toku15,toku171,toku17}, it is important to employ two types of projection-operator methods to derive the basic equations for different relevant variables. In fact, it is indispensable to use the time-convolutionless formalism for the number densities to recover the cumulant expansion proposed by Kubo \cite{kubo62}. A linear non-Markov stochastic diffusion equation for $\rho(\bm{q},t)$ is then derived from the Heisenberg equation by employing the Tokuyama-Mori projection-operator method \cite{toku75,toku76} (see a dotted arrow (T) in Fig. \ref{evo}), where the memory terms are convolutionless in time and are written in terms of correlation function of the fluctuating current. A linear non-Markov Langevin type equation for $j(\bm{q},t)$ is also derived from the Heisenberg equation by using the Mori projection-operator method \cite{mori65} (see a dashed arrow (M) in Fig. \ref{evo}), where the memory term is convolution in time and is written in terms of correlation function of the fluctuating forces. By introducing the cumulant function $K(q,t)$ by $K(q,t)=-\ln[f(q,t)]$, those coupled equations are then used to obtain the following closed nonlinear non-Markov second-order differential equation for $K(q,t)$ (see a bold T arrow in Fig. \ref{evo}): \cite{toku171}
\begin{eqnarray}
\frac{\partial^2}{\partial t^2}K(q,t)=\frac{q^2v_{th}^2}{S(q)}&-&\zeta_0\frac{\partial}{\partial t}K(q,t)\nonumber\\
&-&\int_0^tds\int_0^sd\tau\Delta\varphi(q,s-\tau)\frac{f(q,\tau)}{f(q,s)}\frac{\partial^2}{\partial\tau^2}K(q,\tau)
\label{Keq1}
\end{eqnarray}
with the nonlinear memory function
\begin{equation}
\Delta\varphi(q,t)=\frac{\rho v_{th}^2}{2}\int_<\frac{d\bm{k}}{(2\pi)^3}v(\bm{q},\bm{k})^2S(k)S(|\bm{q}-\bm{k}|)f(k,t)f(|\bm{q}-\bm{k}|,t),\label{memory}
\end{equation}
where $\zeta_0$ is a friction coefficient, $\int_<$ the sum over wave vectors $\bm{k}$ whose magnitudes are smaller than a cutoff $q_c$ \cite{toku19}, and $v_{th}=(k_BT/m)^{1/2}$. The vertex amplitude $v(\bm{q},\bm{k})$ is given by 
\begin{equation}
v(\bm{q},\bm{k})=\hat{\bm{q}}\cdot\bm{k}c(k)+\hat{\bm{q}}\cdot(\bm{q}-\bm{k})c(|\bm{q}-\bm{k}|),\label{vertex}
\end{equation}
where $\rho c(k)=1-1/S(k)$ and $\hat{\bm{q}}=\bm{q}/q$. This is a TMCT equation to discuss the dynamics of glass-forming equilibrium liquids. Here we note that the nonlinear memory function $\Delta\varphi(q,t)$ has exactly the same form as that obtained in MCT \cite{mct84L,mct84B,mct91}, although TMCT equation is quite different from MCT equation. By using the Percus-Yevic static structure factor \cite{py}, one can solve this equation numerically and show that it describes not only the $\alpha$-relaxation process but also the $\beta$-relaxation process consistently \cite{toku171}. However, we should mention here that because of the double time integrals in the memory function, such numerical calculations are limited only for smaller volume fractions far from the glass transition. In general, therefore, it is difficult to obtain numerical solutions for higher values of $\xi$ near the glass transition. In a slow stage [K$_2$], however, one can avoid such a difficulty because one can apply the approximation (A2) given by $f(q,\tau)/f(q,s)\simeq 1$ \cite{toku171} to Eq. (\ref{Keq1}) (see a upper bold arrow in Fig. \ref{evo}). Thus, one can find the asymptotic second-order differential equation for $K(q,t)$
\begin{equation}
\frac{\partial^2 K(q,t)}{\partial t^2}=\frac{q^2v_{th}^2}{S(q)}-\zeta_0\frac{\partial K(q,t)}{\partial t}
-\int_0^t\Delta\varphi(q,t-s)\frac{\partial K(q,s)}{\partial s}ds, \label{ka}
\end{equation}
where the initial conditions are given by $K_{\alpha}(q,t=0)=dK_{\alpha}(q,t)/dt|_{t=0}=0$. This is an ideal TMCT equation to discuss the dynamics of glass-forming equilibrium liquids. Here we note that because of the approximation (A2), the numerical solutions of Eq. (\ref{ka}) coincide with those of Eq. (\ref{Keq1}), except at $\beta$ stage.

In a hydrodynamic stage [H],  the diffusion equation for $K(q,t)$ is derived from either Eq. (\ref{Keq1}) or Eq. (\ref{ka})  in the long time limit as (see a lower bold arrow in Fig. \ref{evo})
\begin{equation}
 \frac{\partial K(q,t)}{\partial t}=q^2D_c(q),  \;\;\text{or}\;\; \frac{\partial f(q,t)}{\partial t}=-q^2D_c(q)f(q,t) \label{deq}
\end{equation}
with the $q$-dependent collective diffusion coefficient
\begin{equation}
D_c(q)=\frac{v_{th}^2/S(q)}{\zeta_0+\int_0^{\infty}\Delta\varphi(q,s)ds}. \label{dcf} 
\end{equation}

\subsection{an ergodic to nonergodic transition}
We now discuss an ergodic to non-ergodic transition at a critical point $\xi=\xi_c$, above which the long-time solution $f(q,t)$ (or $K(q,t)$) reduces to a non-zero value $f_c(q)$ (or $K_c(q)$), the so-called nonergodicity parameter. As shown in the previous papers \cite{toku15,toku171,toku17}, starting from either Eq. (\ref{Keq1}) or Eq. (\ref{ka}), in the long-time $t\rightarrow \infty$ limit, one can find 
\begin{equation}
K_c(q)=\frac{1}{\mathpzc{F}(q)} \label{ene}
\end{equation}
with the long-time limit of the nonlinear memory function
\begin{equation}
\mathpzc{F}(q,f_c,f_c)=\frac{1}{2}\int_< \frac{d\bm{k}}{(2\pi)^3}\Theta(\bm{q},\bm{k})f_c(k)f_c(|\bm{q}-\bm{k}|),\label{memo22}
\end{equation}
where the vertex $\Theta$ is given by
\begin{equation}
\Theta(\bm{q},\bm{k})=(\rho/q^2) v(\bm{q},\bm{k})^2S(q)S(k)S(|\bm{q}-\bm{k}|).\label{Vm}
\end{equation}
The existence of the critical point in Eq. (\ref{ene}) is mathematically confirmed since $K_c(q)$ is a kind of the Lambert W-function. In fact, the critical volume fraction $\phi_c$ has been obtained by solving numerically Eq. (\ref{ene}) with the Percus-Yevick static structure factor as $\phi_c\simeq 0.5817$ \cite{kim16,naru17,toku19}, which agrees with that obtained from the simulation results performed on monodisperse hard spheres \cite{toku03,toku031,toku05,toku071} within error. In the long-time limit we obtain $K(q,t)=q^2D_c(q)t$ for $\xi<\xi_c$, while we find $K(q,t)=K_c(q)$ and $D_c(q)=0$ for $\xi\geq \xi_c$ since the integral term of the memory function $\Delta\varphi(q,s)$ in Eq. (\ref{dcf}) diverges at the critical point $\xi_c$. Thus, the existence of the nonergodicity parameter suggests a singularity as a function of the distance $(1-\xi/\xi_c)$ at $\xi=\xi_c$. In fact, by employing the same mathematical approach as that used in MCT \cite{mct84B,mct91}, one can also write $D_c(q)$ in terms of a singular function as 
\begin{equation}
D_c(q,\xi)\propto (1-\xi/\xi_c)^{\gamma}, \label{ltdq}
\end{equation}
where $\gamma$ is a power exponent to be determined.

\subsection{Characteristic decays}
We briefly review two types of characteristic decays near the glass transition, which are used in this paper. One is the so-called von Schweidler (VS) decay \cite{vsc} at the slow $\beta$ stage given by
\begin{equation}
f(q,t)\simeq f_c(q)[1-(t/t_b)^b], \label{vsd}
\end{equation}
 where $b$ is a time exponent and $t_b$ a characteristic time. Another is the so-called stretched exponential decay (or KWW function) \cite{kw,ww} at $\alpha$ stage given by 
\begin{equation}
f(q,t)=f_c(q)\exp[-(t/t_{\alpha})^{\beta}], \label{KWW}
\end{equation}
where $\beta$ is the KWW exponent and $t_{\alpha}$ the $\alpha$-relaxation time. As shown in Ref. \cite{toku191}, near the critical point $\xi_c$, the crossover time $t_x$ from the VS decay to the KWW decay is given by $t_x\simeq K_c^{1/\beta}t_{\alpha}\simeq K_c^{1/b}t_b$. By using Eq. (\ref{ka}), one can also find a simple relation between the exponent parameter $\lambda$ \cite{mct91}, $b$, and $\beta$ as \cite{toku191}
\begin{equation}
\lambda=\frac{\Gamma[1+b]^2}{\Gamma[1+2b]}=\frac{\Gamma[2\beta-1]}{\Gamma[2\beta+1]\Gamma[\beta+1]\Gamma[-\beta-1]}, \label{expbeqla}
\end{equation}
where $\Gamma[x]$ is an usual $\Gamma$-function. In the following, these results are used to discuss how the new model is different from the conventional model not only qualitatively but also quantitatively.

\subsection{Universal TMCT equation}
In this subsection, we discuss the universal equation to describe the dynamics of glass-forming liquids of type $i$ with the intensive control parameters such as an inverse temperature, where $i=$(F) for fragile liquids and $i=$(S) for strong liquids. As shown in the previous papers \cite{toku07,toku09,toku10,toku11,toku13,toku141,toku16,toku131}, the mean-$n$th displacement in a liquid of type $i$ coincides with the other mean-$n$th displacements in different liquids of type $i$ if the universal parameter $u_s$ in those liquids has the same value, where $u_s(q=0)=-\log(D_s(q=0)q_m/v_{th})$. Here $D_s(q=0)$ is a long-time self-diffusion coefficient. However, we note that the displacement in (F) never coincides with that in (S), even if $u_s$ has the same value. In general, these situations also hold for the cumulant function $K(t)(=K(q_m,t))$. We discuss this next.

Since the characteristic features between different liquids of type $i$ can be compared at a first peak of the static structure factor, we set $q=q_m$ in Eq. (\ref{ka}). Then, Eq. (\ref{ka}) depends on the physical quantities $q_m$, $v_{th}$, and $S(q_m)$. In order to eliminate them from Eq. (\ref{ka}), it is convenient to introduce the time scale $\tau_D$ as
\begin{equation}
\tau_D=S(q_m)^{1/2}/(q_mv_{th}). \label{tauD}
\end{equation}
By using a scaled time $\tau=t/\tau_D$, one can then find the following asymptotic solution of Eq. (\ref{ka}):
\begin{equation}
K(\tau)\simeq\begin{cases} \tau^2/2, & \text{$\tau\ll 1$},\\
\tau/\tau_L, & \text{$\tau>\tau_{\alpha}$},
\end{cases} \label{assol}
\end{equation}
with the diffusion time $\tau_L(=D^{-1})$, where $D(q_m)$ is the scaled diffusion coefficient given by
\begin{equation}
D(q_m)=q_m^2\tau_D D_c(q_m)=q_mS(q_m)^{1/2}D_c(q_m)/v_{th}. \label{sltdc}
\end{equation}
For a short time the ballistic motion dominates the dynamics of the system, while for a long time the diffusion process dominates it. For such time regions there are no difference between the dynamics of (F) and that of (S). The clear difference between them appears at $\alpha$ and $\beta$ stages through the nonlinear memory function $\Delta\varphi(q_m,\tau)$. One can then transform Eq. (\ref{ka}) into a dimensionless equation
\begin{equation}
\frac{\partial^2 K(\tau)}{\partial \tau^2}=1-\zeta\frac{\partial K(\tau)}{\partial \tau}
-\kappa\int_0^t M(\tau-s)\frac{\partial K(s)}{\partial s}ds, \label{kasc}
\end{equation}
where $\zeta=\zeta_0\tau_D$ and $M(\tau)=\Delta\varphi(q_m,\tau)/\Delta\varphi(q_m,0)$. Here the coupling parameter $\kappa$ is given by $\kappa=\tau_D^2\Delta\varphi(q_m,0)$. This is an universal equation to describe the dynamic of glass-forming equilibrium liquids of type $i$.

In order to solve Eq. (\ref{kasc}) numerically, one needs to fix the values of two unknown parameters $\zeta$ and $\kappa$ consistently. As shown in the previous papers \cite{toku15,toku17}, this is done by using the simulation results at each value of $D(T)$. In fact, the friction coefficient $\zeta_0$ has been found to be constant from the short-time behavior of the simulation results \cite{toku17}. Since $\zeta$ depends on $T$ through $\tau_D$, one can fix the value of $\zeta$ at each value of $D$. The coupling parameter $\kappa$ is also found by using the simulation results under the fixed value of $D$. In fact, from Eqs. (\ref{dcf}) and (\ref{sltdc}), one can obtain
\begin{equation}
\kappa=\frac{1- \zeta D}{D\int_0^{\infty}M(\tau) d\tau}. \label{kappa}
\end{equation}
Thus, the coupled equations (\ref{kasc}) and (\ref{kappa}) can be solved self-consistently at a given value of $D$.
Hence the universal parameter $u$ for $K(\tau)$ is now given by
\begin{equation}
u=-\log(D(T)). \label{up}
\end{equation}
Later, we also show that the scaled $\alpha$-relaxation time $\tau_{\alpha}$ is related to $D$ as $\tau_{\alpha}\propto D^{-1}$. 

\subsection{Recursion equation convenient for a simplified memory function}
In the previous paper \cite{toku17}, we have fixed a value of $\varepsilon$ so that the plateau height of the numerical solution coincides with that of the simulation results. However, this approach is not appropriate because the numerical solutions of Eq. (\ref{ka}) with the PY static structure factor are known to deviate from the simulation results at $\beta$ stage \cite{toku19,toku171}. In order to solve Eq. (\ref{kasc}) with the simplified memory function, therefore, we start from the simulation results as an initial condition. Then, it is convenient to use the following formal solution of Eq. (\ref{kasc}):
\begin{equation}
K(\tau)=\frac{K_0(\tau)-\int_0^{\tau} \Sigma(s)[K(\tau-s)-K(\tau)]ds}{1+\int_0^{\tau} \Sigma(s)ds} \label{Ktssc}
\end{equation}
with 
\begin{equation}
\Sigma(s)=\kappa\int_0^s e^{-\zeta(s-s')}M(s')ds', \label{Gs}
\end{equation}
 where $K_0(\tau)=(\zeta\tau-1+e^{-\zeta\tau})/\zeta^2$. This is a starting equation to find the numerical solutions by iterations. The nonlinear exponents $\mu$ and $\varepsilon$ are then found so that starting from the simulation results, the first iteration result coincides with the plateau height of the simulation results around the crossover time $\tau_x$. Several iterations are thus done at the fixed values of $\mu$ and $\varepsilon$ until $\kappa$ reaches a constant value within error of order $10^{-3}$. In the following, we thus obtain the numerical solutions not only for the new model but also for the conventional one and compare them together with the simulation results from a unified point of view consistently. Finally, we note that Eq. (\ref{Ktssc}) is an exact equation derived from Eq. (\ref{kasc}) and is different from the approximate equation derived in the previous paper \cite{toku17}, where $K(\tau-s)$ has been expanded in powers of $s/\tau$, up to order $(s/\tau)^2$.

\section{A new simplified model}
The numerical solutions of Eq. (\ref{Ktssc}) for the conventional model are shown to fail in describing the $\alpha$-relaxation process near the glass transition. This means that the integration over the wavevector $\bm{k}$ in the nonlinear memory function $\Delta\varphi(q_m,t)$ must play an important role in finding a reasonable value of $\beta$ at $\alpha$ stage. In order to perform the integration approximately, one may simply assume that $S(k)$ obeys a Gaussian distribution with the peak position $q_m$ as $S(k)\simeq a_0\exp[-a_1(k-q_m)^2]$, where $a_0$ and $a_1$ are positive constants. We also assume around $k=q_m$ that $K(k,\tau)\simeq (k/q_m)^2K(q_m,\tau)$ and $K(|\bm{q}_m-\bm{k}|,\tau)\simeq K(q_m,\tau)$. Then, the integration of Eq. (\ref{memory}) over $\bm{k}$ leads to an asymptotic function of $(1+\mu K)^{-1}$, where $\mu$ is a fitting parameter to be determined. Thus, we propose the following simplified nonlinear memory function:
\begin{equation}
M(\tau)=\frac{f(\tau)}{[1+\mu K(\tau)]^{\nu}}, \label{asmf}
\end{equation}
where $\nu$ is a fitting parameter to be determined. 

This model contains two unknown parameters $\mu$ and $\nu$. Since $\mu K(\tau)=-\ln(f(\tau)^{\mu})$, the parameter $\mu$ must play an role of a nonlinear exponent in $f(\tau)$. On the other hand, the parameter $\nu$ must be found through the dynamics of $\alpha$- and $\beta$-relaxation processes near the glass transition. In fact, at a slow $\beta$ stage, $f(\tau)$ is described by the von Schweidler decay given by Eq. (\ref{vsd}) (or $K(\tau)\simeq K_c+(\tau/\tau_b)^b$). As discussed in Ref. \cite{toku191}, the VS decay dominates the system, up to the crossover time $\tau_x$. In $\alpha$ stage for $\tau\geq \tau_x$, therefore, the memory function $M(s)$ in the time integral of Eq. (\ref{Ktssc}) is dominated by the VS decay $K(s)\sim s^b$. From Eq. (\ref{asmf}), we then obtain $M(s) \sim f_c s^{-\nu b}$. On the time scale of order $\tau(\geq \tau_x\geq s)$, one can expand $K(\tau-s)$ in powers of $s/\tau$ as $K(\tau-s)\simeq K(\tau)+O(s/\tau)$. Thus, Eq. (\ref{Ktssc}) reduces to
\begin{equation}
K(\tau)\sim \frac{\tau}{\int_0^{\tau} s^{-\nu b}ds}\sim \tau^{\nu b}. \label{ltasK}
\end{equation}
This must be compared with the KWW function given by $K(\tau)\sim (\tau/\tau_{\alpha})^{\beta}$ to find
\begin{equation}
\nu=\beta/b. \label{nure}
\end{equation}
Here the numerical value of $\nu$ is found by using the numerical values of $\beta$ and $b$ obtained in Ref. \cite{toku191}, which are listed in Table \ref{table-1}.
\begin{table}
\caption{Time exponents $b$ and $\beta$, exponent parameter $\lambda$, and exponent $\nu$ for different systems \cite{toku191}.}
\begin{center}
\begin{tabular}{ccccc}
\hline
System  & $b$ & $\beta$ & $\lambda$& $\nu(=\beta/b)$ \\
\hline
control parameter $1/T$ &&&&\\
Fragile liquids &  0.3064 & 0.6832 & 0.8980& 2.230\\
Strong liquids &  0.2779 & 0.6809 &0.9130& 2.451\\
\hline
control parameter $\phi$ &&&&\\
Hard spheres & 0.6051 & 0.710 &0.7215&1.173\\
Soft spheres  &0.6051 & 0.710& 0.7215&1.173\\
\hline
\end{tabular}
\end{center}
\label{table-1}
\end{table}

\begin{figure}
\begin{center}
\begin{tabular}{cc}
\begin{minipage}{.5\textwidth}
\includegraphics[width=6.7cm]{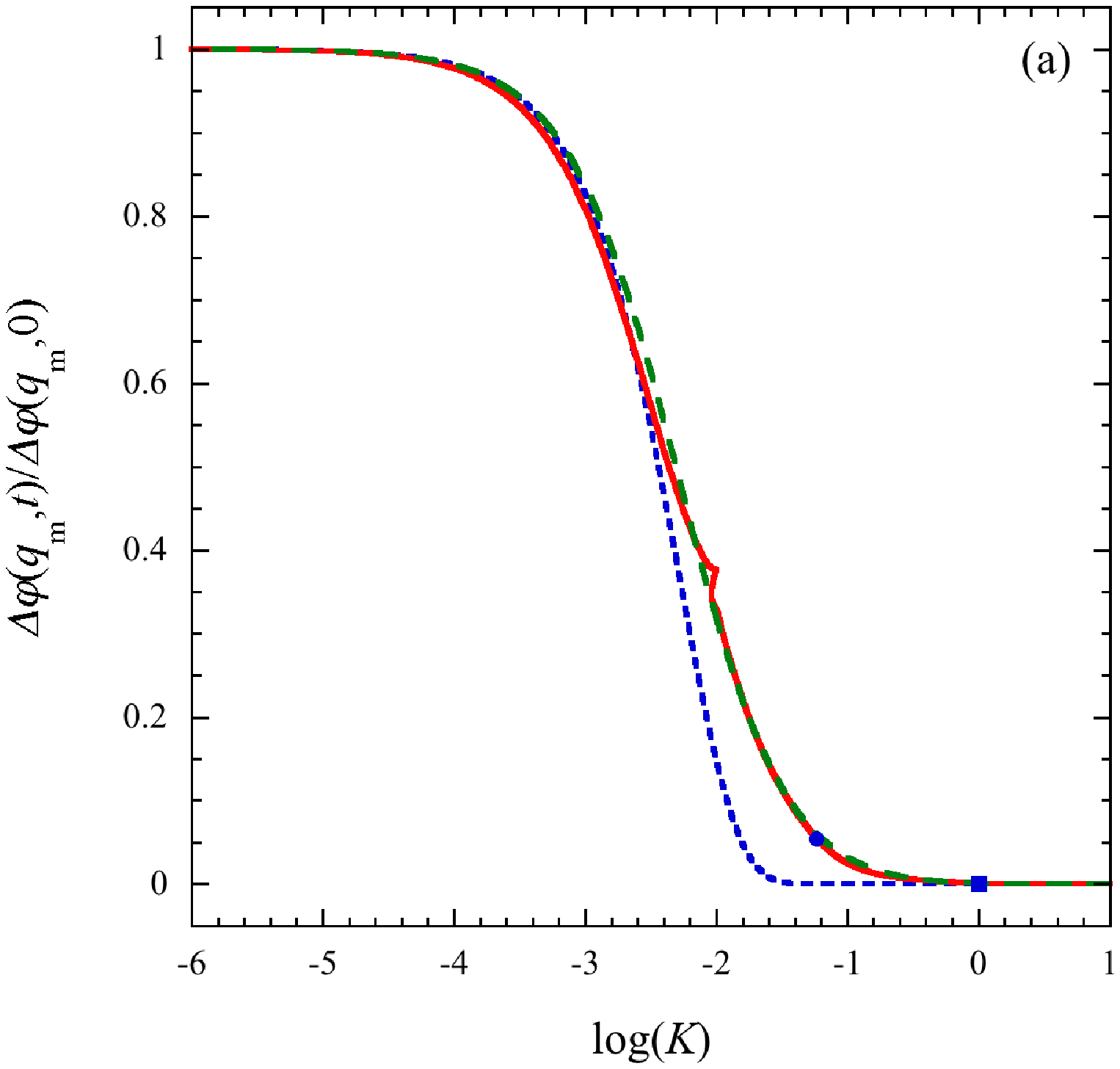}
\end{minipage}
\begin{minipage}{.5\textwidth}
\includegraphics[width=6.7cm]{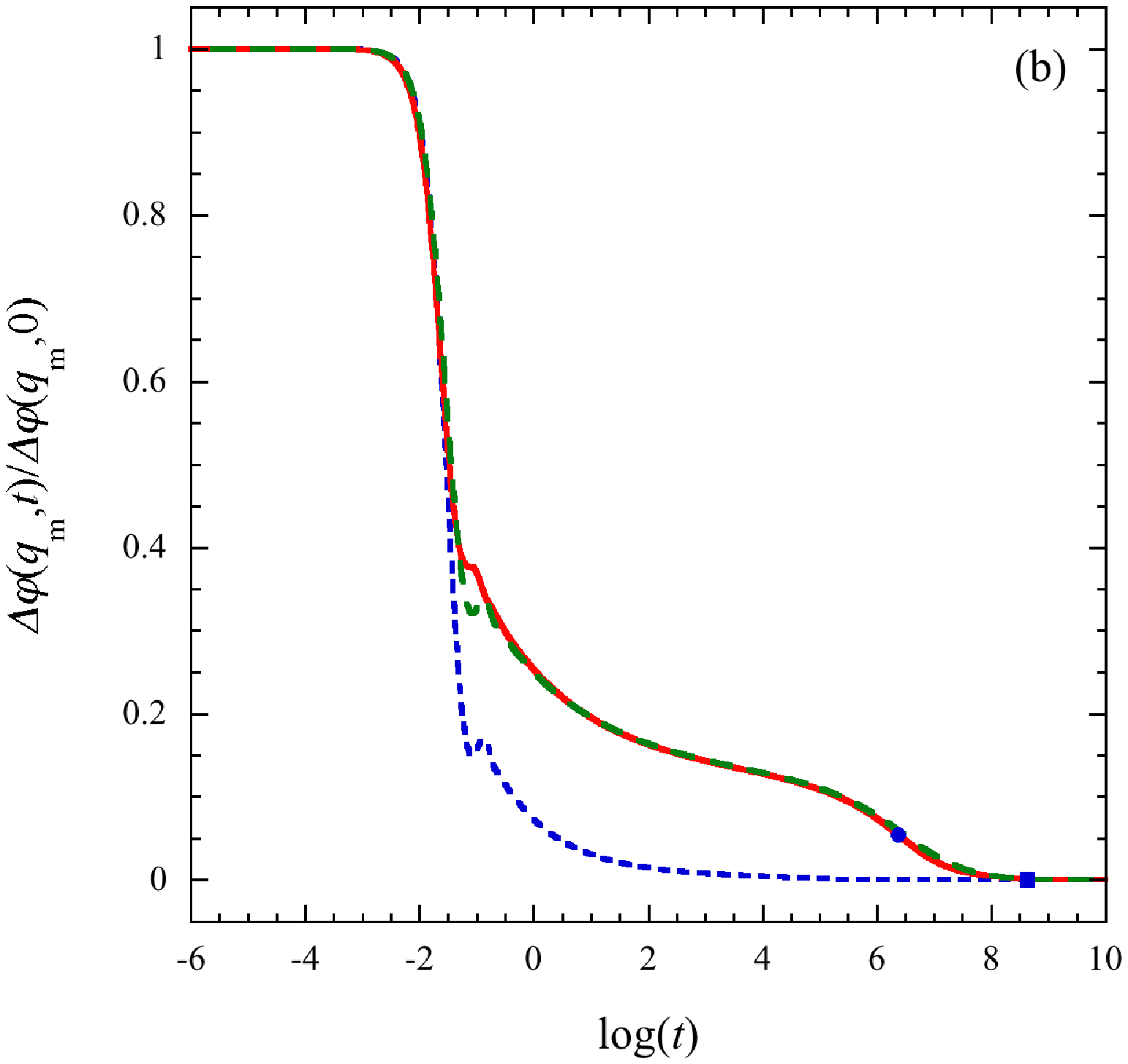}
\end{minipage}
\end{tabular}
\end{center}
\caption{(Color online) A plot of the scaled memory function $\Delta\varphi(q_m,t)/\Delta\varphi(q_m,0)$ versus (a) $\log(K)$ and (b) $\log(t)$. The solid line indicates the scaled memory function obtained from the numerical calculation of Eq. (\ref{ka}) at $\phi=0.581$, $q_m\sigma=7.4$, and $q_c\sigma=60$ \cite{narup}. The long-dashed line indicates the numerical values of Eq. (\ref{asmf}) at $\nu=1.173$ and $\mu=162.21$. The dashed line indicates the numerical value of Eq. (\ref{rsmf}) at $\varepsilon=191.27$. The symbols $(\bullet)$ indicate the crossover time $t_x=10^{6.375}$ ($K_x=10^{-1.233}$) and $(\Box)$ the $\alpha$-relaxation time $t_{\alpha}=10^{8.631}$ ($K_{\alpha}=1$).}
\label{memK}
\end{figure}
In order to check whether the new model is reasonable or not, one may use the scaled memory function $\Delta\varphi(q_m,t)/\Delta\varphi(q_m,0)$ directly obtained from the numerical calculation of the ideal TMCT equation given by Eq. (\ref{ka}) with the PY static structure factor \cite{narup}. In Fig. \ref{memK}, the calculated scaled memory function is plotted versus (a) $\log(K)$ and (b) $\log(t)$ together with the new model given by Eq. (\ref{asmf}) and the conventional model given by Eq. (\ref{rsmf}). The calculated memory function is then shown to be well described by the new model at $\nu=1.173$ and $\mu=162.21$ within error. Thus, the new model is expected to describe the simulation results well not only qualitatively but also quantitatively. On the other hand, the conventional model can describe the calculated memory function only for a short time ($t<10^{-2}$) and a long time ($t>t_{\alpha}$). In fact, one can write Eq. (\ref{asmf}) as $M(t)\simeq \exp[-(1+\mu\nu)K]$ for a short time. From Eq. (\ref{rsmf}), one obtains $\varepsilon=1+\mu\nu$. The conventional model is then shown to coincide with the calculated memory function only for a short and a long times at $\varepsilon=191.27$ (see Fig. \ref{memK}), where $\nu=1.173$ and $\mu=162.21$. Thus, the conventional model is easily shown to disagree with the calculated memory function at $\alpha$ and $\beta$ stages even for any other values of $\varepsilon$. This is the main reason why it fails to describe the $\alpha$-relaxation process.

\section{Numerical solutions of universal TMCT equation}
In the present section, we solve Eq. (\ref{Ktssc}) based on the simulation results.

\subsection{Simulation results}
We first briefly summarize the simulation results. As shown in the previous papers \cite{toku141,toku16,toku131}, the molecular-dynamics simulations have been performed for two types of glass-forming liquids, (F) fragile liquids and (S) strong liquids, where the control parameter $\xi$ is given by an inverse temperature $1/T$. By using those simulation results, one can then find the physical quantities, such as $q_m$ and $\zeta_0$. As typical examples, we here take the binary mixtures A$_{80}$B$_{20}$ with the Stillinger-Weber potential (SW) \cite{sw} for (F) and SiO$_2$ with the Nakano-Vashishta potential (NV) \cite{nv} for (S). The detailed information about those simulations is found in Refs. \cite{toku141,toku16,toku131}. In SW, length, time, and temperature are scaled by $\sigma$, $t_0(=\sigma/v_0)$, and $\epsilon/k_B$, respectively, where $\epsilon$ is energy unit, and $v_0=(\epsilon/m)^{1/2}$. In NV, length, time, and temperature are measured by \AA, ps, and K, respectively. The particle mass is given by average mass over components. The friction constant $\zeta_0$ is then found from the simulations as $\zeta_0(T)\simeq$ 12 (SW) and 96 (NV), while the peak position $q_m$ of the total static structure factor $S(q)$ is chosen near the glass transition as $q_m\sigma=$7.25 (SW) and 1.55 (NV). 
\begin{table}[b]
\caption{Critical temperature $T_c$, glass transition temperature $T_g$, supercooled temperature $T_s$, deeply supercooled temperature $T_f$, and corresponding universal parameters $u_g=u(T=T_g)$, $u_s=u(T=T_s)$, and $u_f=u(T=T_f)$ for different systems.}
\begin{center}
\begin{tabular}{ccccccccc}
\hline
type & system & $T_c$ & $T_g$ &$T_f$ &$T_s$&$u_g$&$u_f$&$u_s$\\
\hline
fragile &SW&0.505 & 0.527 &0.644 &0.891&6.082&3.102&1.705\\

\hline
strong &NV &2490.4  &2826.5 &3317.9 &3830.0 &4.09
0&2.620&1.920\\
\hline
\end{tabular}
\end{center}
\label{table-2}
\end{table}

\begin{figure}
\begin{center}
\begin{tabular}{cc}
\begin{minipage}{.5\textwidth}
\includegraphics[width=6.7cm]{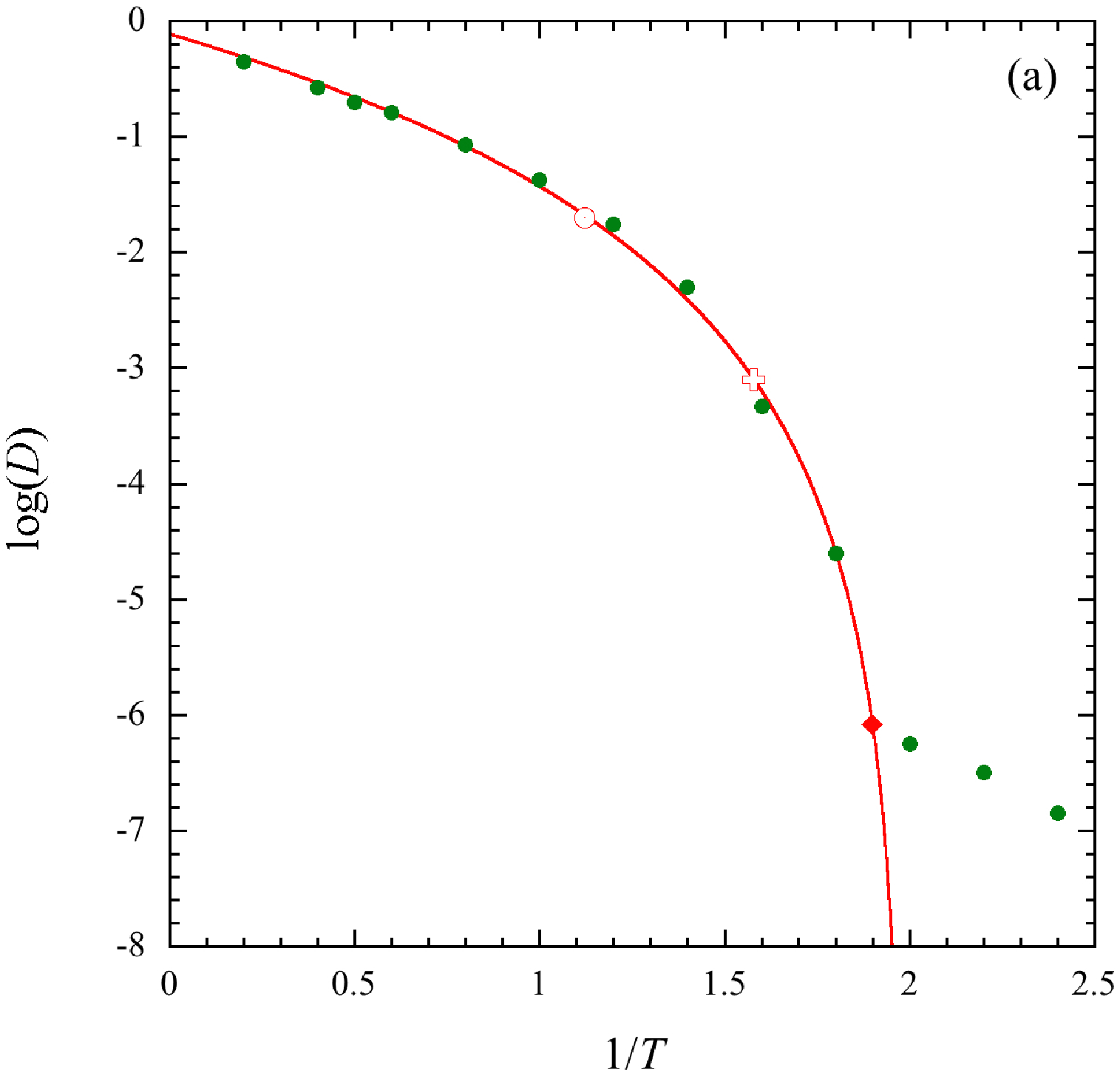}
\end{minipage}
\begin{minipage}{.5\textwidth}
\includegraphics[width=6.7cm]{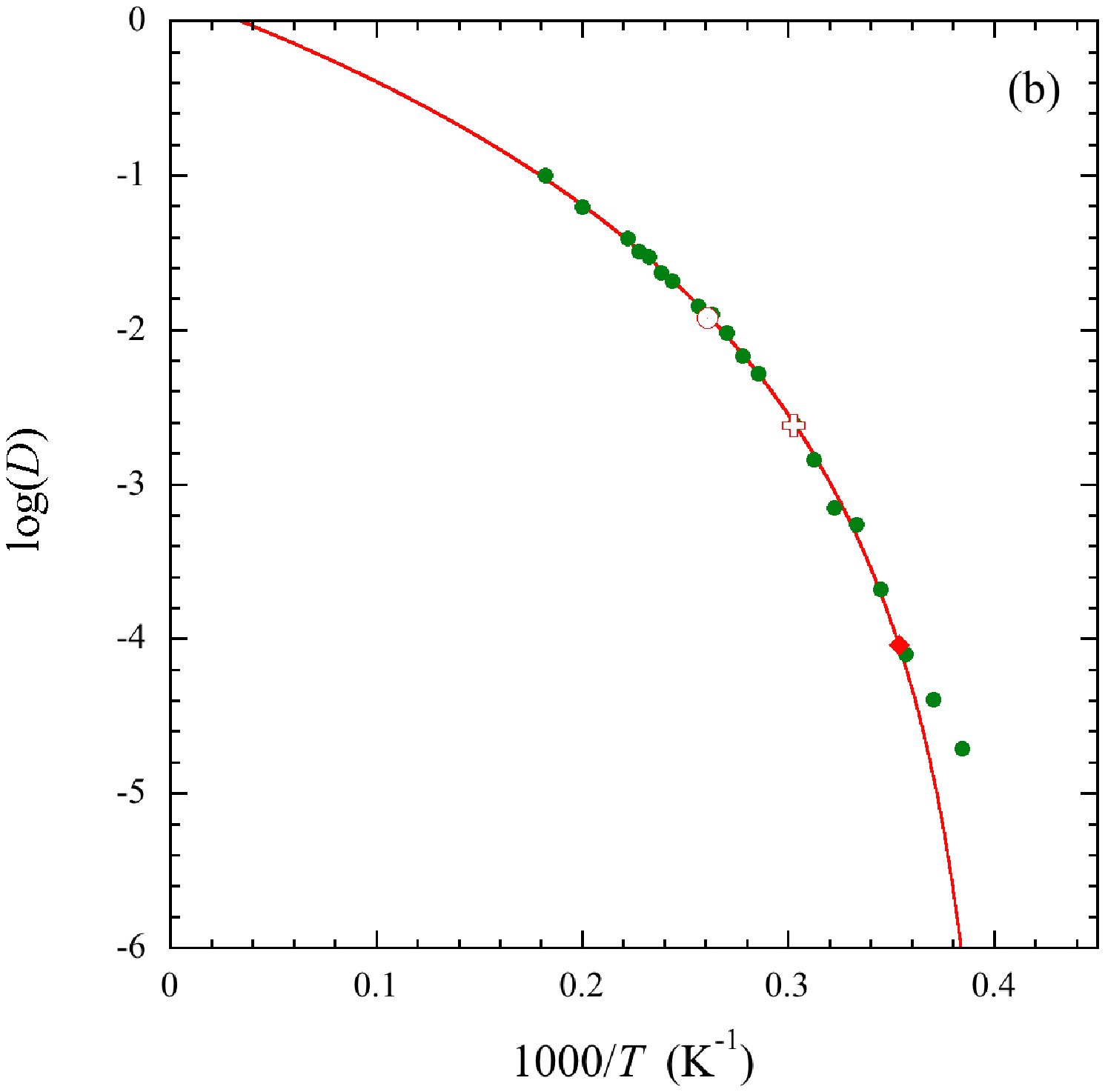}
\end{minipage}
\end{tabular}
\end{center}
\caption{(Color online) A log plot of the scaled diffusion coefficient $D$ versus inverse temperature for (a) a fragile liquid and (b) a strong liquid. The symbols ($\bullet$) indicate the simulation results. The solid lines indicate the singular function given by Eq. (\ref{sltd}). The symbols ($\Diamond$) indicate the glass transition point $T_g$, ($\odot$) the supercooled point $T_s$, and ($+$) the deeply supercooled point $T_f$, whose numerical values are listed in Table \ref{table-2}.}
\label{da}
\end{figure}
The numerical values of $D$ and $\zeta$ are found from the simulation results. In Fig. \ref{da}, the simulation results for $D(q_m)$ are then plotted versus inverse temperature together with the singular function of $D$. As discussed in Refs. \cite{toku17,toku13}, the value of the power exponent $\gamma$ of $D_c$ depends on type $i$. In fact, we have $\gamma=4.0$ for (F) and 4.333 for (S) \cite{toku13}. Similarly to Eq. (\ref{ltdq}), one may also assume that $D$ obeys the singular function near $T_c$ as
\begin{equation}
D(T)=A_r(1-T_c/T)^{\gamma_r}, \label{sltd}
\end{equation}
where $\gamma_r$ is a power exponent to be determined and $A_r$ a prefactor to be determined. From the simulation results, we thus find $\gamma_r\simeq 4.317$ and $A_r\simeq 0.775$ for (F) and $\gamma_r\simeq 4.563$ and $A_r\simeq 1.500$ for (S). Here we note that $\gamma_r$ is slightly different from $\gamma$ because $D$ also depends on $S(q_m)$ and $v_{th}$ through Eq. (\ref{sltdc}). By using the simulation results, one can also find $\tau_{\alpha}$ in terms of $D$ as 
\begin{equation}
\tau_{\alpha}(T)=B_r/D(T)=B_r\tau_L(T), \label{taud}
\end{equation}
where $B_r\simeq 0.822$ for (F) and 0.875 for (S). Hence we obtain $\tau_L/\tau_{\alpha}\simeq 1.214$ for (F) and 1.182 for (S).

\begin{figure}
\begin{center}
\begin{tabular}{cc}
\begin{minipage}{.5\textwidth}
\includegraphics[width=6.7cm]{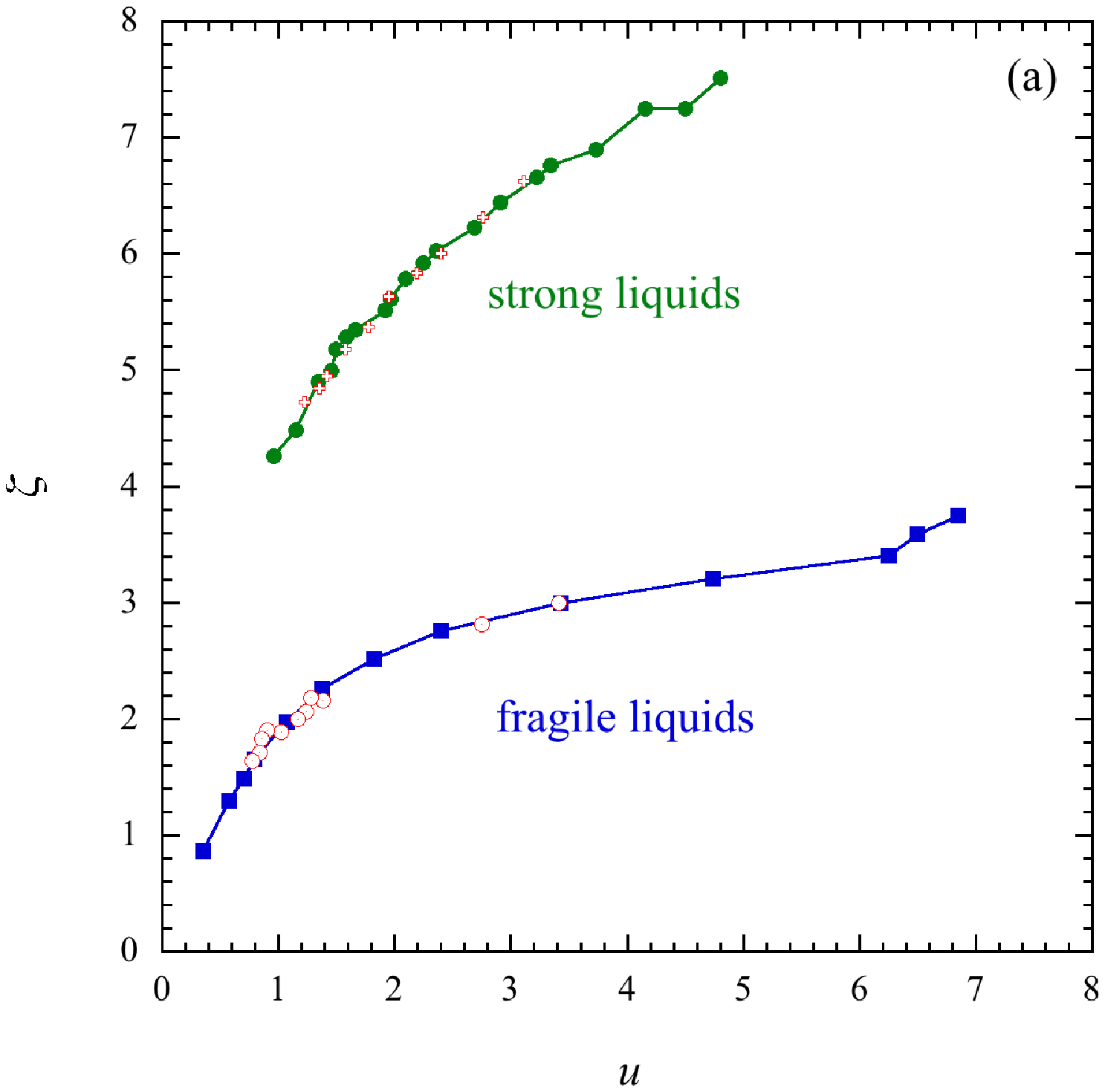}
\end{minipage}
\begin{minipage}{.5\textwidth}
\includegraphics[width=6.7cm]{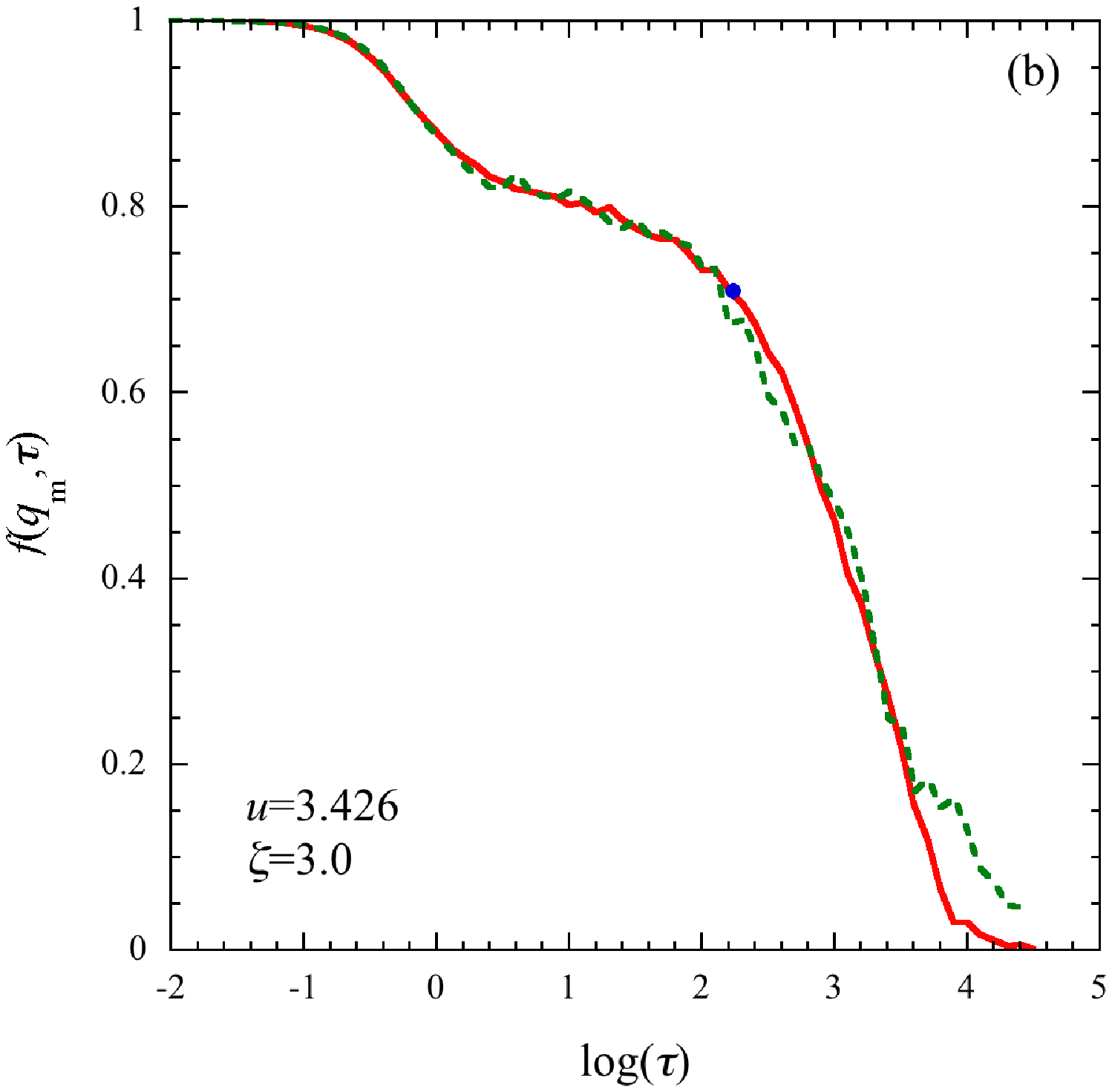}
\end{minipage}
\end{tabular}
\end{center}
\caption{(Color online) Universal function of $u$. (a) A plot of the scaled friction coefficient $\zeta$ versus $u$. The symbols ($\Box$) indicate the numerical values for SW, $(\odot$) for Al$_2$O$_3$, ($\bullet$) for NV, and ($+$) for BKS. The solid lines are guides for eyes. (b) A plot of $f(q_m,\tau)$ versus $\log(\tau)$ at $u\simeq 3.426$ and $\zeta\simeq 3.0$. The solid line indicates the simulation results for SW at $T=0.625$ and the dashed line for Al$_2$O$_3$ at $T=2300$ (K) \cite{toku17}.}
\label{uni}
\end{figure}
\begin{figure}
\begin{center}
\begin{tabular}{cc}
\begin{minipage}{.5\textwidth}
\includegraphics[width=6.7cm]{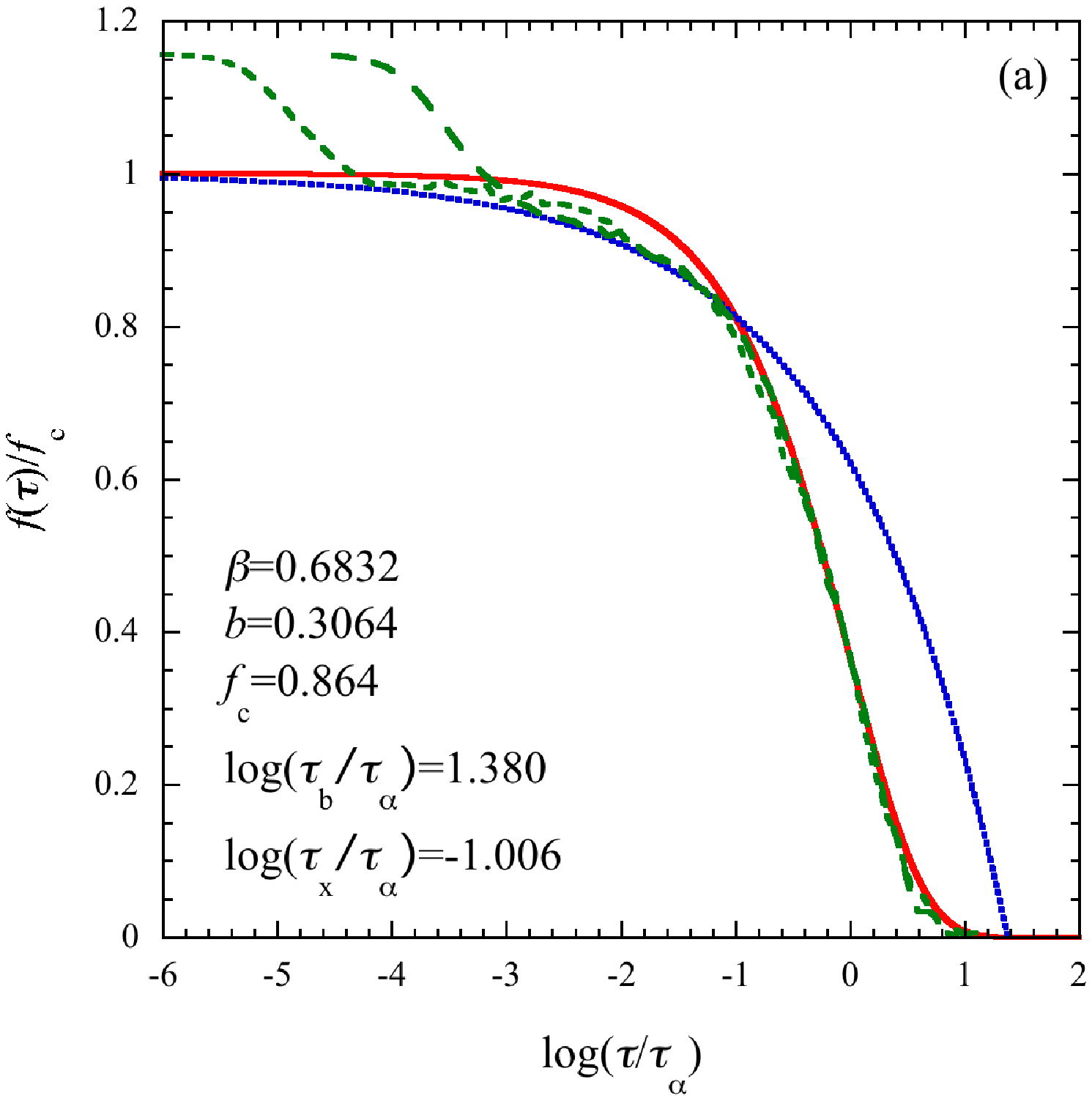}
\end{minipage}
\begin{minipage}{.5\textwidth}
\includegraphics[width=6.7cm]{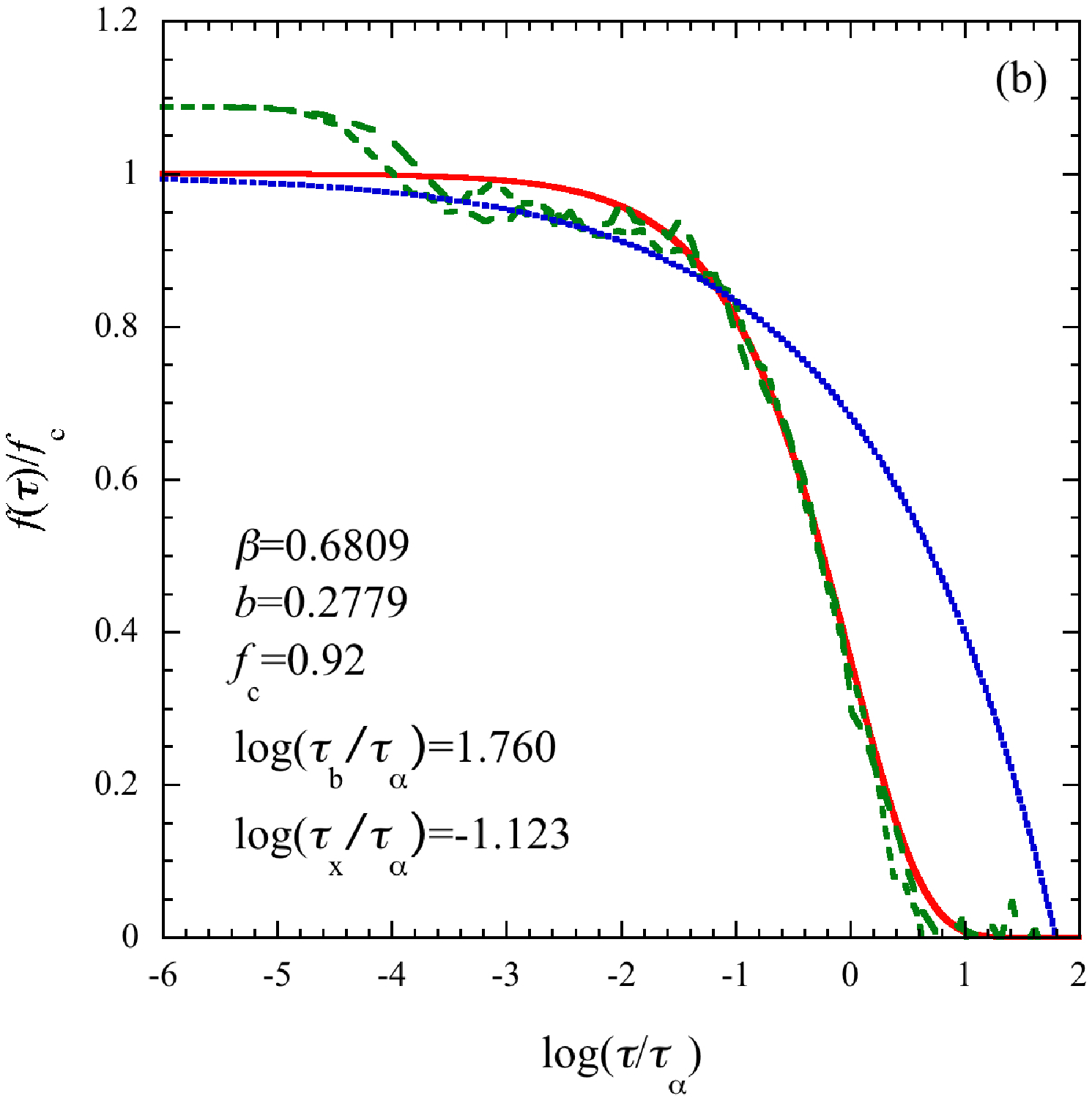}
\end{minipage}
\end{tabular}
\end{center}
\caption{(Color online) A plot of the scaled function $f(\tau)/f_c$ versus $\log(\tau/\tau_{\alpha})$ near the glass transition. (a) The long-dashed and the dashed lines indicate the simulation results for SW at $T=$0.625 and 0.556, respectively, where $f_c=0.864$. Here $\log(\tau_{\alpha})=3.312$ for $T=0.625$ and 4.621 for 0.556. (b) The long-dashed and the dashed lines indicate the simulation results for NV at $T=$2900 and 2800 (K), respectively, where $f_c=0.920$. Here $\log(\tau_{\alpha})=3.674$ at $T=2900$ (K) and 4.101 at 2800 (K). In both figures, the solid lines indicate the scaled KWW decay and the dotted lines the scaled VS decay.}
\label{stsim}
\end{figure}
In Fig. \ref{uni}(a), the scaled friction coefficient $\zeta$ is plotted versus $u$. Although $\zeta_0$ is constant, $\zeta$ increases as $u$ increases because it depends on $S(q_m)$ and $v_{th}$. Here we note that $\zeta$ is an universal function of $u$ in different systems of type $i$. In fact, the values of $\zeta$ obtained from the simulation results \cite{toku17} for Al$_2$O$_3$ with the Born-Meyer potential \cite{bm} and for SiO$_2$ with the Beest-Kramer-Santen (BKS) potential \cite{bks} are also shown in Fig. \ref{uni}(a), where $\zeta_0\simeq$ 82 and $q_m\sigma$=4.25 for Al$_2$O$_3$ and $\zeta_0\simeq$ 113 and $q_m\sigma$=1.65 for BKS. In each type of liquids $\zeta$ is thus shown to coincide with each other. Hence the dynamics in different systems of type $i$ is expected to coincide with each other if $u$ has the same value in those systems \cite{toku17}. As one of such examples, in Fig. \ref{uni}(b) the simulation results for SW at $T=0.625$ are compared with those for Al$_2$O$_3$ at $T=2300$K, where $u\simeq 3.426$ and $\zeta\simeq 3.0$ in both systems. Both simulation results are thus shown to coincide with each other within error. Here we note that those are only available data to be compared at the same value of $u$ in [S]. This is because those simulations have been done independently. In order to confirm the universality in dynamics, therefore, the simulations in various systems should be done at the same value of $u$.

As discussed in Ref. \cite{toku191}, the simulation results satisfy the relation given by Eq. (\ref{expbeqla}) near the glass transition, where the relevant exponents are listed in Table \ref{table-1}. In Fig. \ref{stsim}, we show how the simulation results for SW and NV are well described by the VS decay and the KWW decay near the glass transition within error. Those results are also used later to check from a unified point of view wether the numerical solutions for both models can be described by those decays well or not.

\subsection{Numerical solutions}
\begin{figure}
\begin{center}
\begin{tabular}{cc}
\begin{minipage}{.5\textwidth}
\includegraphics[width=6.5cm]{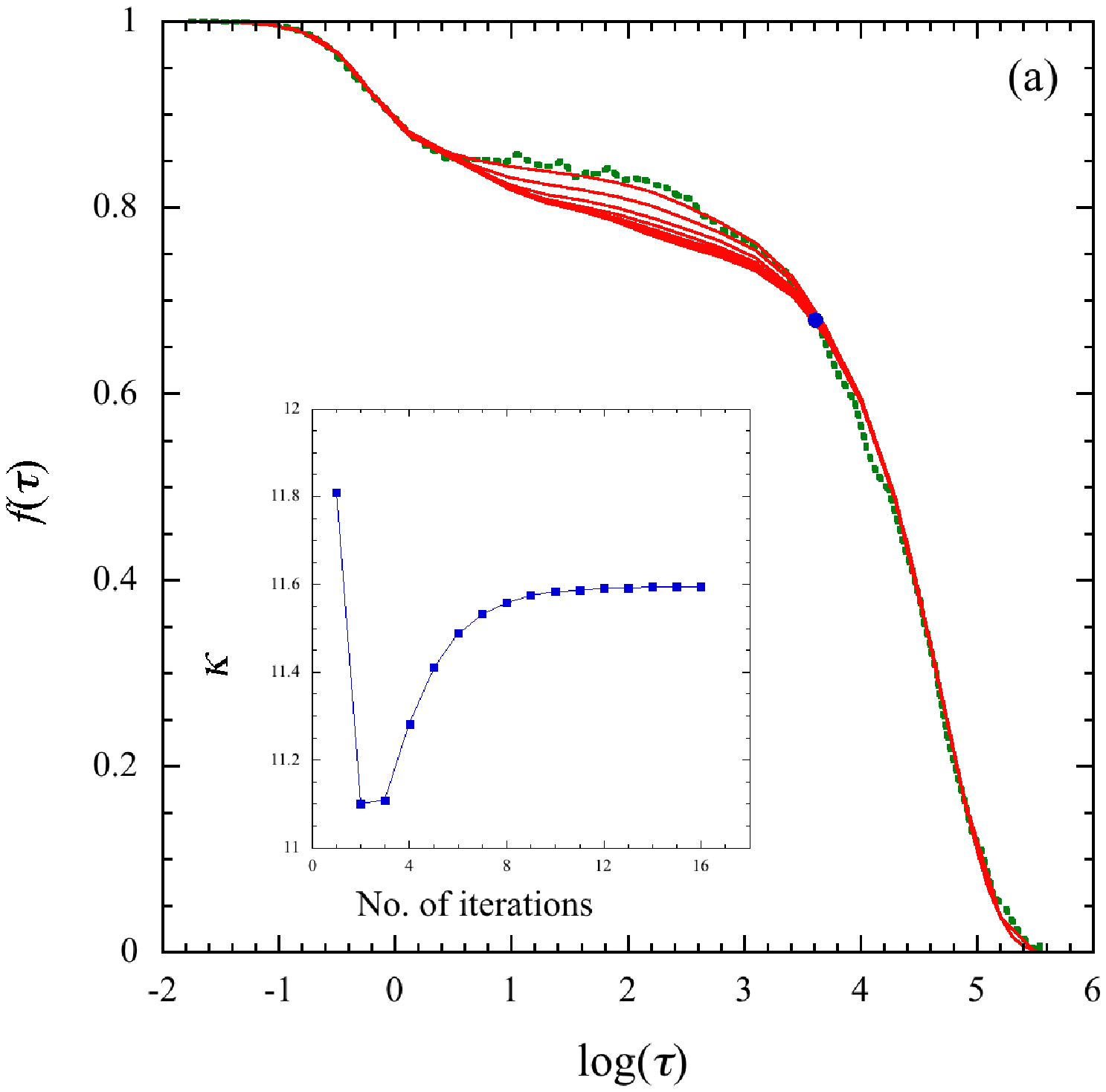}
\end{minipage}
\begin{minipage}{.5\textwidth}
\includegraphics[width=6.5cm]{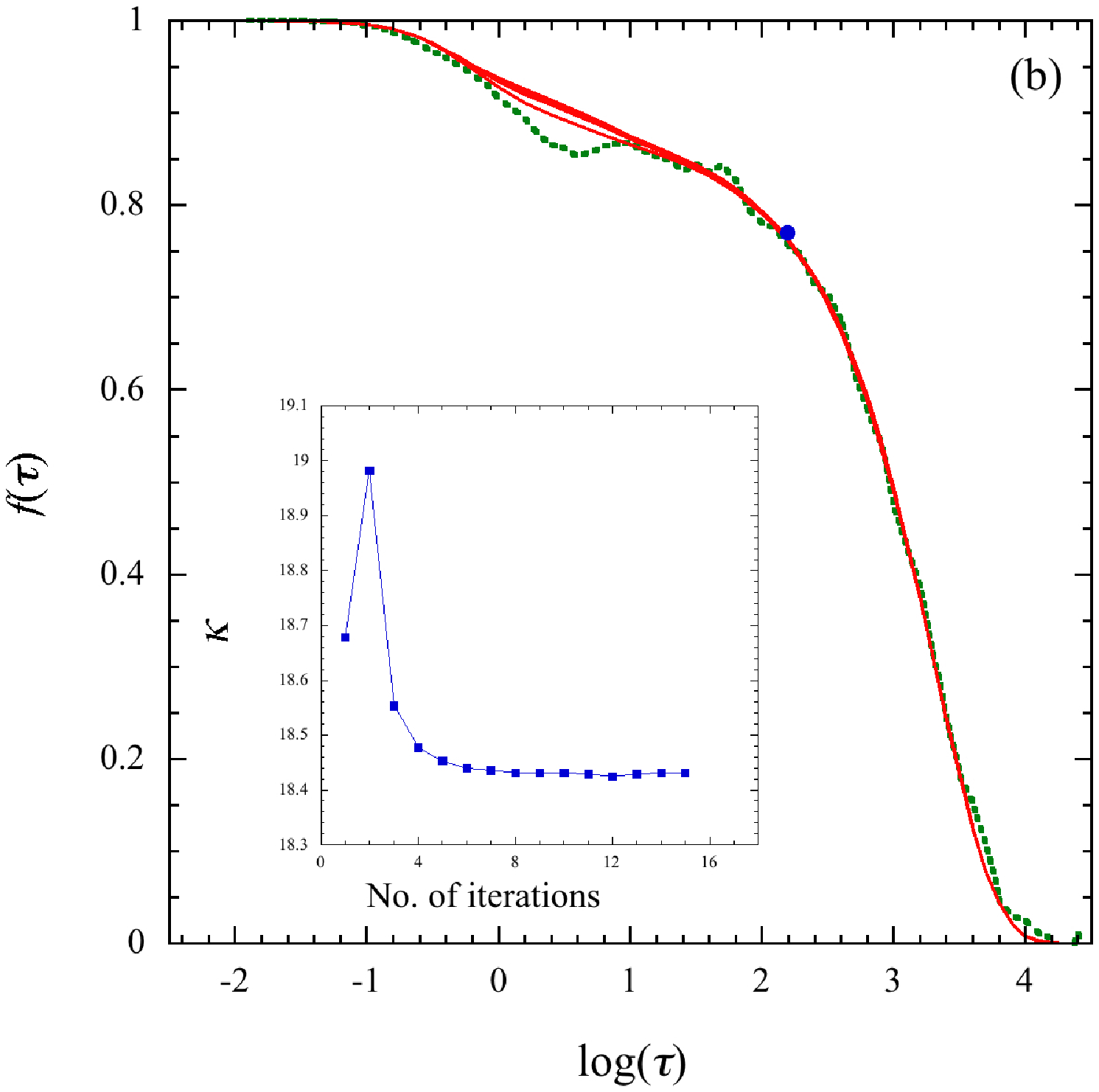}
\end{minipage}
\end{tabular}
\end{center}
\caption{(Color online) A log plot of the scattering function $f(\tau)$ versus $\tau$ near the glass transition. (a) The dotted line indicates the simulation results for SW at $T=0.556$ $(u\simeq 4.740)$. The solid lines indicate the numerical solutions for the new model obtained by 16th iterations (from top to bottom) at $\mu=1.8$. The symbol $(\bullet)$ indicates the crossover time $\log(\tau_x)\simeq 3.615$. (b) The dotted line indicates the simulation results for NV at $T=3000$ (K) $(u\simeq 3.340)$. The solid lines indicate the numerical solutions for the new model obtained by 15th iterations (from bottom to top) at $\mu=3.2$. The symbol $(\bullet)$ indicates the crossover time $\log(\tau_x)\simeq 2.192$. The symbols ($\Box$) in the insets indicate the numerical values of $\kappa$ obtained by iterations. The solid lines in the insets are guides for eyes.}
\label{snit}
\end{figure}

We now solve Eq. (\ref{Ktssc}) for the new model by iterations numerically.
In Figs. \ref{snit}(a) and \ref{snit}(b), the iteration results are plotted together with the simulation results near the glass transition for SW and NV, respectively. Starting from the simulation results, the numerical solutions are shown to deviate from the simulation results only at $\beta$ stage for $1.0<\tau\leq\tau_x$. As the iteration increases, the plateau height decreases in SW, while it increases in NV. This difference is clearly seen in $\kappa$. As the iteration increases, $\kappa$ increases in SW and reaches a constant value around 11.595 within error of order $10^{-3}$ after 16th iterations, while it decreases in NV, reaching a constant value around 18.430 within error of order $10^{-3}$ after 15th iterations. In general, the number of iterations increases as $T$ decreases. 
\begin{figure}
\begin{center}
\begin{tabular}{cc}
\begin{minipage}{.5\textwidth}
\includegraphics[width=6.7cm]{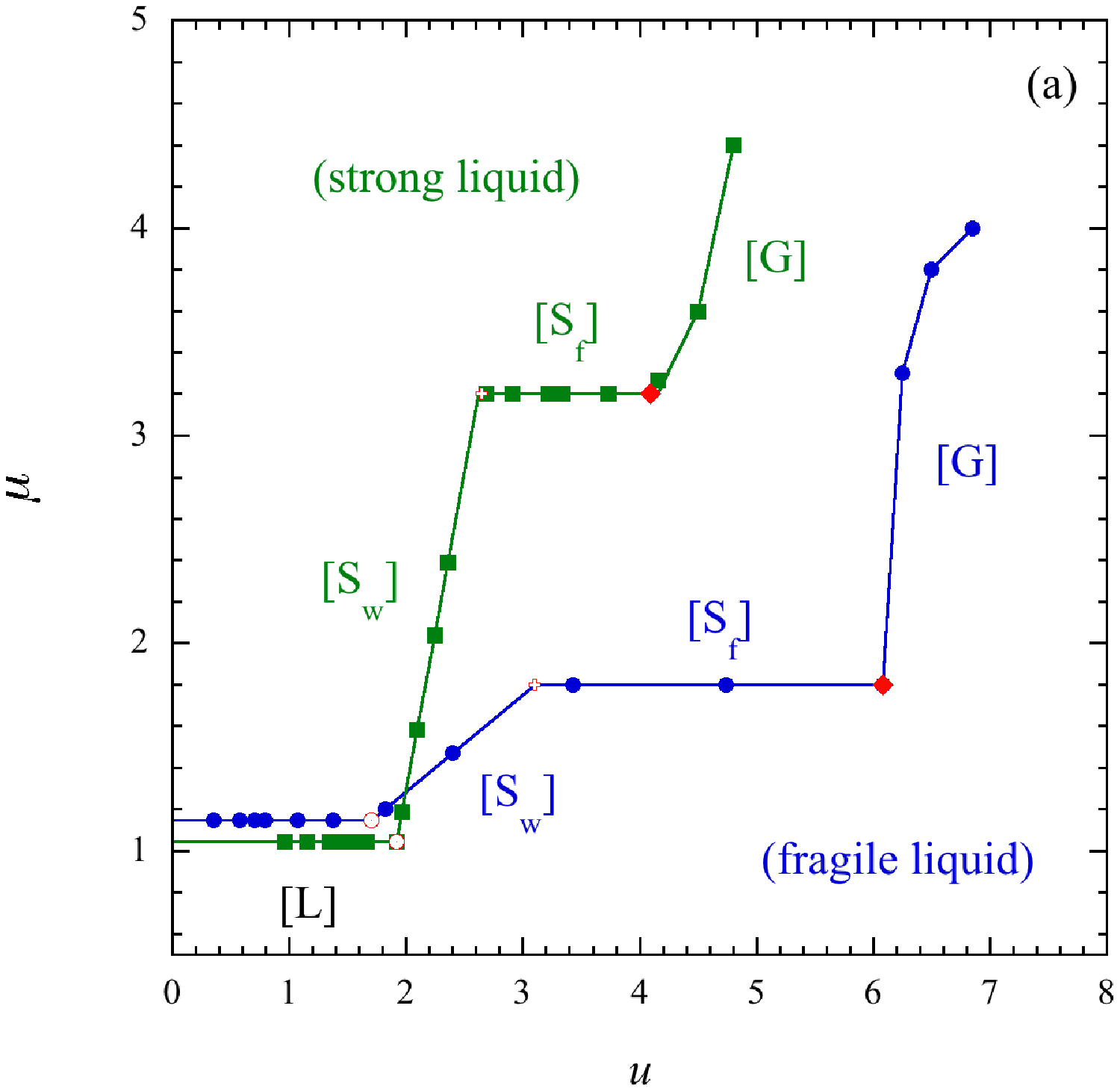}
\end{minipage}
\begin{minipage}{.5\textwidth}
\includegraphics[width=6.7cm]{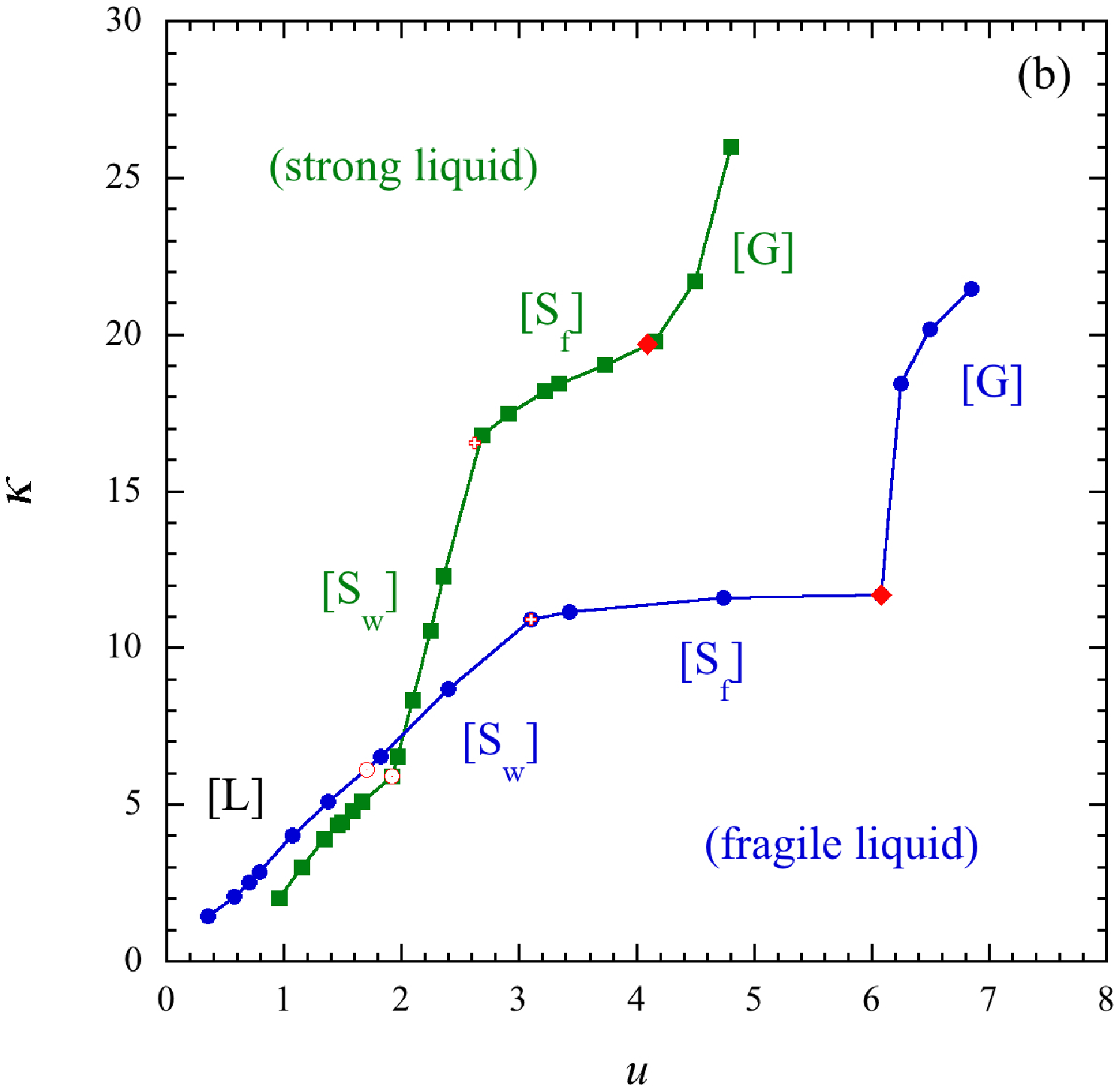}
\end{minipage}
\end{tabular}
\end{center}
\caption{(Color online) A plot of the relevant parameters for the new model versus $u$, where (a) $\mu$ and (b) $\kappa$. In both figures, the symbols ($\bullet$) indicate the fitting values for SW and ($\Box$) for NV. The symbols ($\Diamond$) indicate the glass transition point $u_g$, ($\odot$) the supercooled point $u_s$, and ($+$) the deeply supercooled point $u_f$, where the numerical values of $u_i$ are listed in Table \ref{table-2}. The solid lines are guides for eyes.  The label [L] stands for a liquid state, [S$_w$] for a weakly supercooled state, [S$_f$] for a deeply supercooled state, and [G] for a glass state.}
\label{pfal}
\end{figure}
\begin{figure}
\begin{center}
\begin{tabular}{cc}
\begin{minipage}{.5\textwidth}
\includegraphics[width=6.7cm]{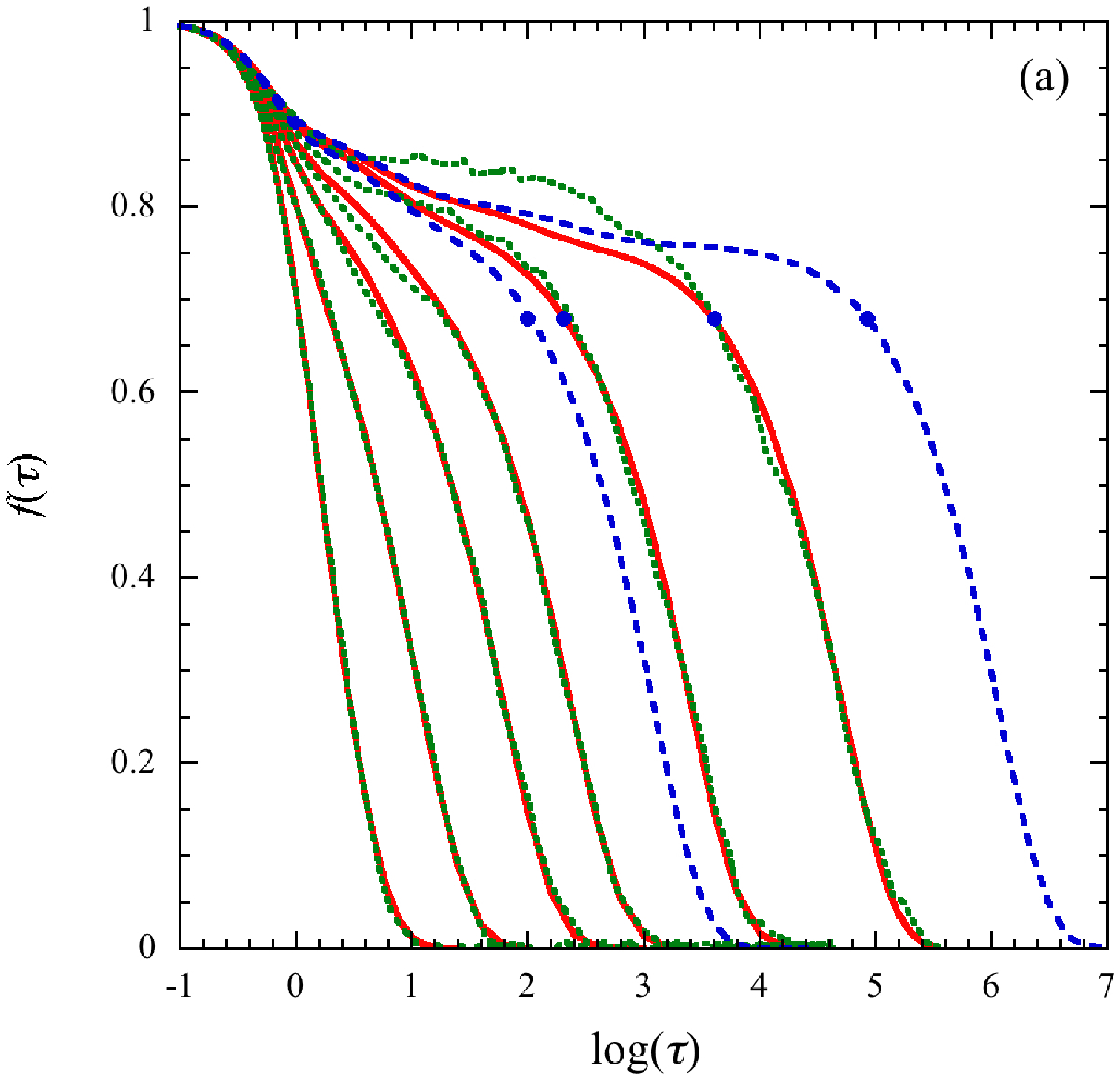}
\end{minipage}
\begin{minipage}{.5\textwidth}
\includegraphics[width=6.7cm]{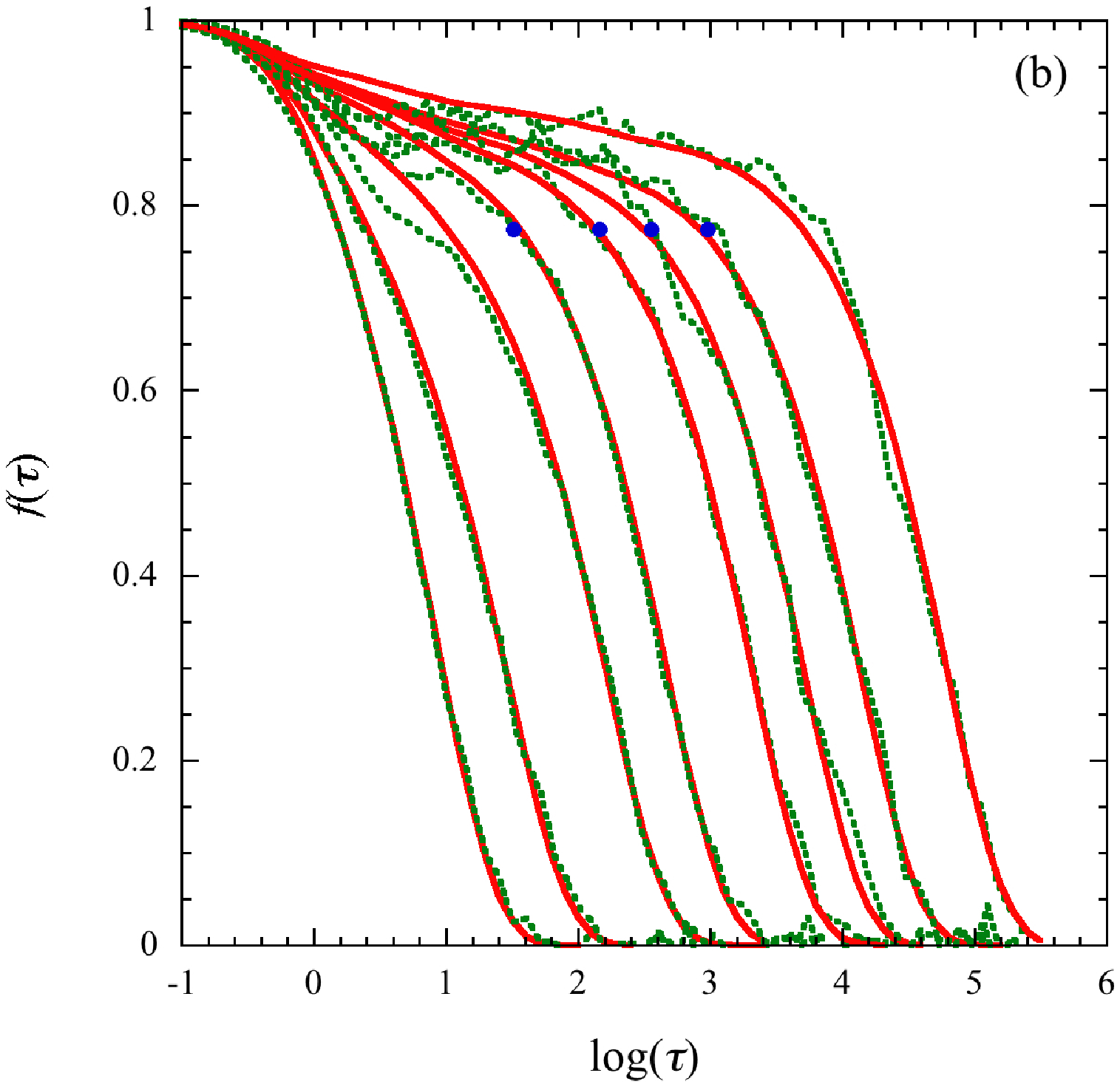}
\end{minipage}
\end{tabular}
\end{center}
\caption{(Color online) A plot of $f(\tau)$ versus $\log(\tau)$ for different temperatures for the new model. (a) The dotted lines indicate the simulation results for SW at $T=$[L] 5.00, 1.25, [S$_w$] 0.833, 0.714, [S$_f$] 0.625, and 0.556 from left to right.  The symbols $(\bullet)$ indicate the crossover time in [S$_f$] given by $\log(\tau_x)\simeq 2.002 ($T=0.644$), 2.306$ (0.625), 3.615 (0.556), and 4.931 (0.527), where $f(\tau_x)\simeq 0.682$. The dashed lines indicate the numerical solutions at $T=$0.644 ($T_f$) and 0.527 ($T_g$). (b) The dotted lines indicate the simulation results for NV at $T=$[L] 5500, 4300, [S$_w$] 3600, [S$_f$] 3300, 3000, 2900, [G] 2800, and 2600 (K) from left to right. The symbols $(\bullet)$ indicate the crossover time in [S$_f$] given by $\log(\tau_x)\simeq 1.562$ ($T=3300$K), 2.192 (3000), 2.551 (2900), and 2.978 (2800), where $f(\tau_x)\simeq 0.775$. The solid lines in both figures indicate the numerical solutions for the new model.}
\label{fall}
\end{figure}
At each temperature $T$ (or $u$) one can thus fix the value of $\mu$ at a first iteration. In Fig. \ref{pfal}(a), the $u$ dependence of the nonlinear parameter $\mu$ is then shown for SW and NV. As $u$ increases, $\mu$ is expected to grow in both systems since the magnitude of the nonlinear fluctuations becomes larger. In fact, the increment of the nonlinearity $\delta\mu(=\mu(u)-\mu(u_s))$ is found in [S]. However, the numerical results for $\mu$ suggest that the supercooled state [S] should be further separated into two substates, a weakly supercooled state [S$_w$] and a deeply supercooled state [S$_f$]. In fact, in each state the value of $\mu$ is given for (F) and (S) by
\begin{equation}
\mu\simeq\begin{cases} 1.150, & \text{[L]\; for $u\leq u_s$}\\
 0.466u+0.355, & \text{[S$_w$] for $u_s\leq u<u_f$}\\
 1.8, & \text{[S$_f$] for $u_f\leq u\leq u_g$} \label{fragile}
\end{cases}
\end{equation}
\begin{equation}
\mu\simeq\begin{cases} 1.046 & \text{[L]\; for $u\leq u_s$}\\
3.077u-4.862, & \text{[S$_w$] for $u_s\leq u<u_f$}\\
3.2, & \text{[S$_f$] for $u_f\leq u\leq u_g$} \label{strong}
\end{cases}
\end{equation}
respectively. In Fig. \ref{pfal}(b), the coupling parameter $\kappa$ is shown to grow rapidly as $u$ increases, except in [S$_f$] where it grows slowly. 

\begin{figure}
\begin{center}
\begin{tabular}{cc}
\begin{minipage}{.5\textwidth}
\includegraphics[width=6.7cm]{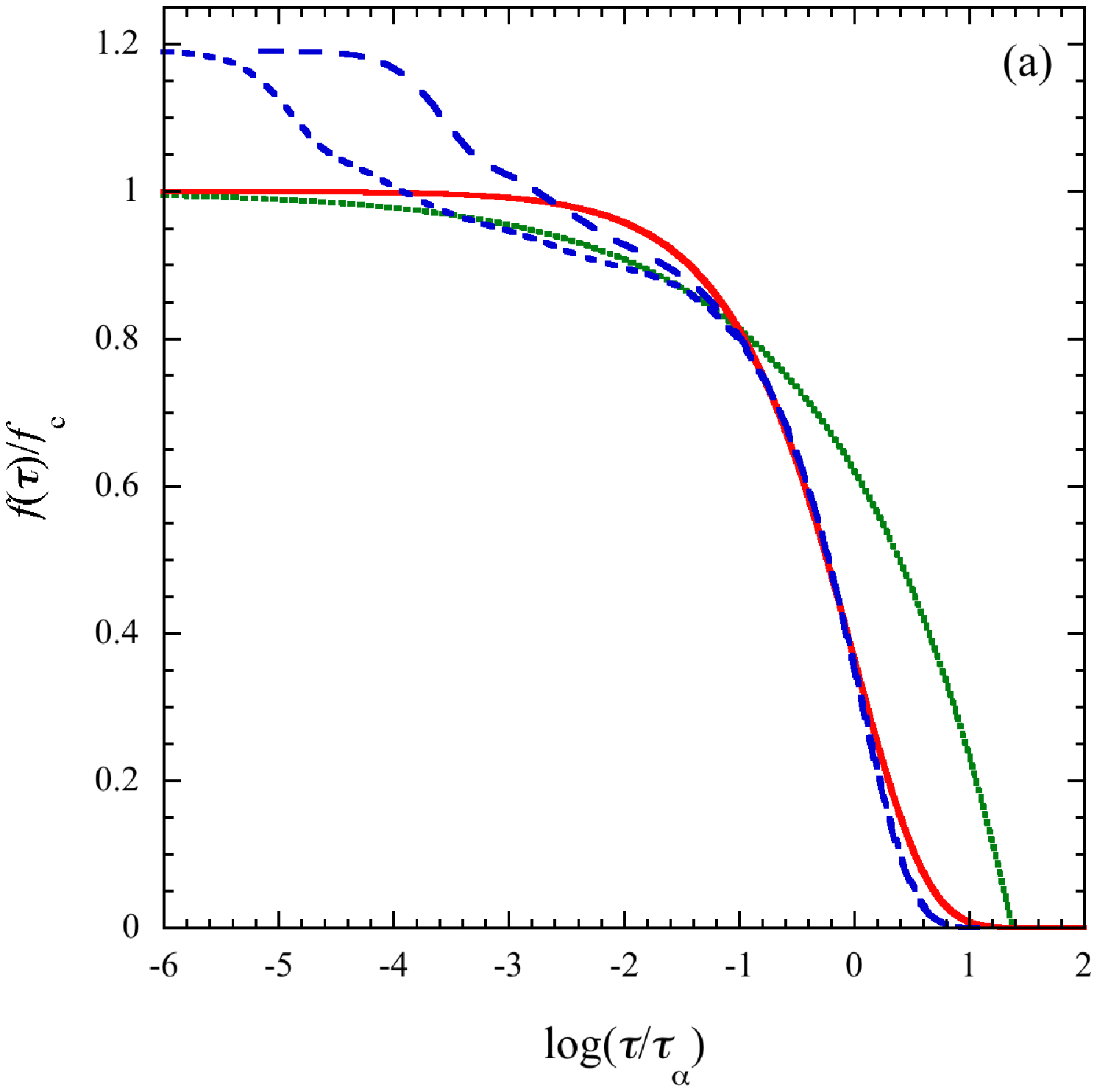}
\end{minipage}
\begin{minipage}{.5\textwidth}
\includegraphics[width=6.7cm]{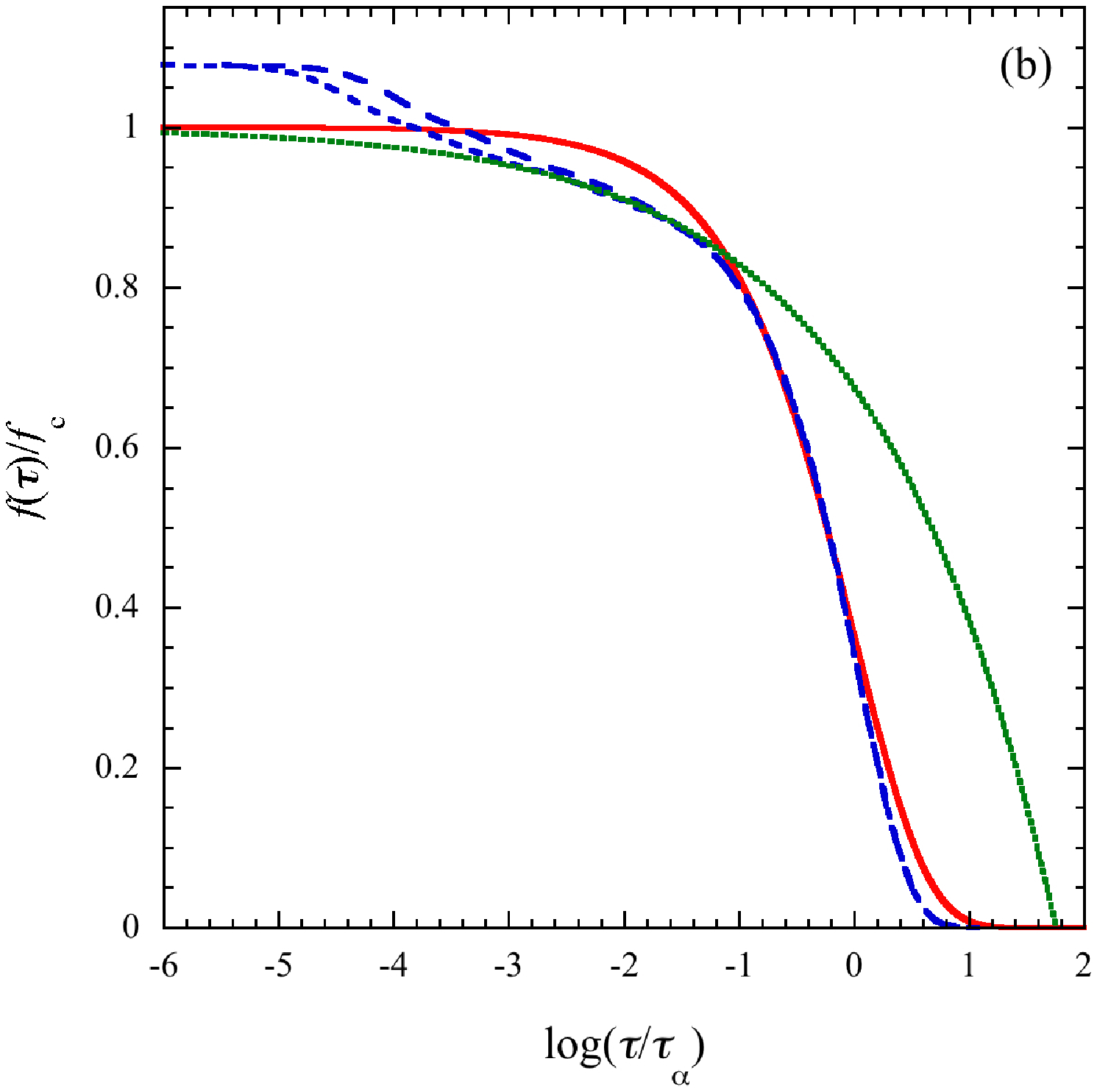}
\end{minipage}
\end{tabular}
\end{center}
\caption{(Color online) A plot of the scaled function $f(\tau)/f_c$ versus $\log(\tau/\tau_{\alpha})$ for the new model near the glass transition. (a) The long-dashed and the dashed lines indicate the numerical solutions for SW at $T=$0.625 and 0.556, respectively, where $f_c=0.840$ and $\log(\tau_b/\tau_{\alpha})=1.380$. (b) The long-dashed and the dashed lines indicate the numerical solutions for NV at $T=$2900 and 2800 (K), respectively, where $f_c=0.928$. The details are the same as in Fig. \ref{stsim}.}
\label{stcal}
\end{figure}
In Figs. \ref{fall}(a) and \ref{fall}(b), the numerical solutions at different temperatures are compared with the simulation results for SW and NV, respectively. For higher temperatures in [L], the solutions are shown to coincide with the simulation results well within error. For lower temperatures, the solutions are shown to agree with them well within error, except at $\beta$ stage. Since one can find a value of $\mu$ approximately in [S], one can obtain the numerical solutions at any temperatures in [S], even though there is no simulation result performed at such temperatures. As typical examples, therefore, we also plot the solutions at $T=T_f$ and $T_g$ in Fig. \ref{fall}(a).  The deviation at $\beta$ stage is clearly seen in a supercooled state [S] and a glass state [G], while they are very small in a liquid state [L]. As pointed out before, such a deviation just results from the ideal TMCT equation. Hence we should mention here that such a technical deviation must disappear if one can solve the original TMCT equation given by Eq. (\ref{Keq1}) for the new model numerically.

Similarly to Fig. \ref{stsim}, we also explore the dynamics near the glass transition carefully by using the KWW function and the VS decay.  The scaled function $f(\tau)/f_c$ is plotted versus scaled time $\tau/\tau_{\alpha}$ in Figs. \ref{stcal}(a) and \ref{stcal}(b) for SW and NV, respectively. Thus, the numerical solutions for the new model in both systems are shown to be described by the KWW function well within error for $\tau_x\leq \tau\leq \tau_{\alpha}$. Here we note that the numerical solutions start to deviate from the KWW function for $\tau>\tau_{\alpha}$. This is just because the long time dynamics is governed by the diffusion process given by $f(\tau)=e^{-\tau/\tau_L}$.

Finally, we compare the numerical results for the new model with those for the conventional model. Starting from the same initial conditions as those for the new model, one can also obtain the numerical solutions for the conventional model consistently. As shown in Fig. \ref{nexT}(b), the $u$ (or $T_c/T$) dependence of $\varepsilon$ is quite different from that of $\mu$. The most important difference appears in [S$_f$]. For a new model, $\mu$ is constant as $T$ decreases, while $\varepsilon$ grows monotonically. On the other hand, in other states the $u$ dependence of both nonlinear parameters is qualitatively similar to each other but quantitatively not. As shown in Ref. \cite{toku17}, the nonlinear parameter $\varepsilon$ also grows linearly in $u$ for (F) and (S). In fact, in each state the value of $\varepsilon$ is given for (F) and (S) by
\begin{equation}
\varepsilon\simeq\begin{cases} 2.564, & \text{[L]\; for $u\leq u_s$}\\
 0.277u+2.092, & \text{[S$_w$] for $u_s\leq u<u_f$}\\
0.152u+2.479, & \text{[S$_f$] for $u_f\leq u\leq u_g$} \label{fragile}
\end{cases}
\end{equation}
\begin{equation}
\varepsilon\simeq\begin{cases} 2.564 & \text{[L]\; for $u\leq u_s$}\\
2.569u-2.357, & \text{[S$_w$] for $u_s\leq u<u_f$}\\
0.376u+3.389, & \text{[S$_f$] for $u_f\leq u\leq u_g$} \label{strong}
\end{cases}
\end{equation}
respectively. The coupling parameter $\kappa$ is also shown to grow monotonically in SW and NV as $u$ increases. The u dependence of $\kappa$ seems to be rather similar to that for the new model. This is just because $\kappa$ is determined by Eq. (\ref{kappa}), where $f(\tau)$ changes, depending on $u$, whether $\mu$ is constant or not in [S$_f$]. 
 
\begin{figure}
\begin{center}
\begin{tabular}{cc}
\begin{minipage}{.5\textwidth}
\includegraphics[width=6.7cm]{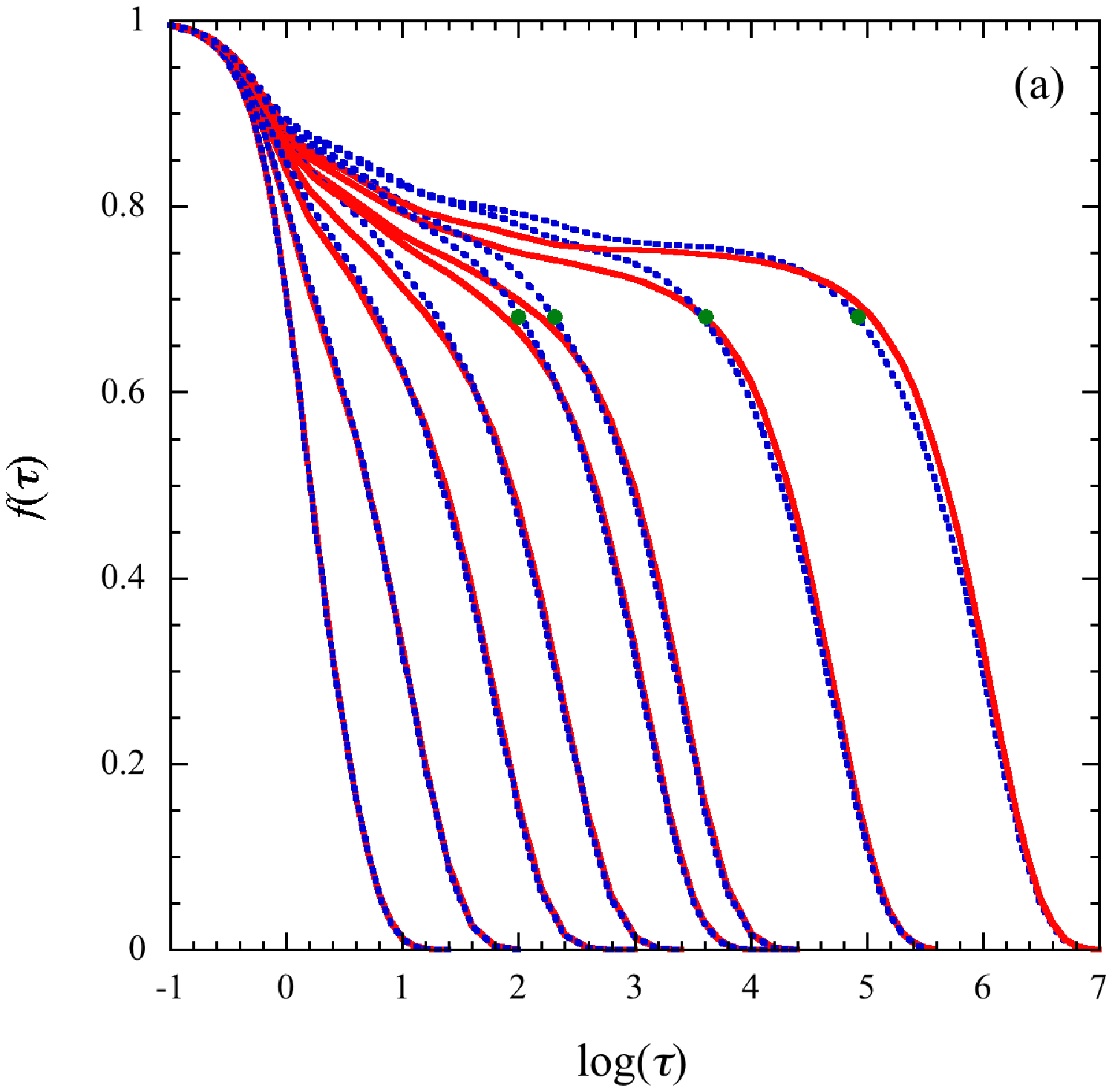}
\end{minipage}
\begin{minipage}{.5\textwidth}
\includegraphics[width=6.7cm]{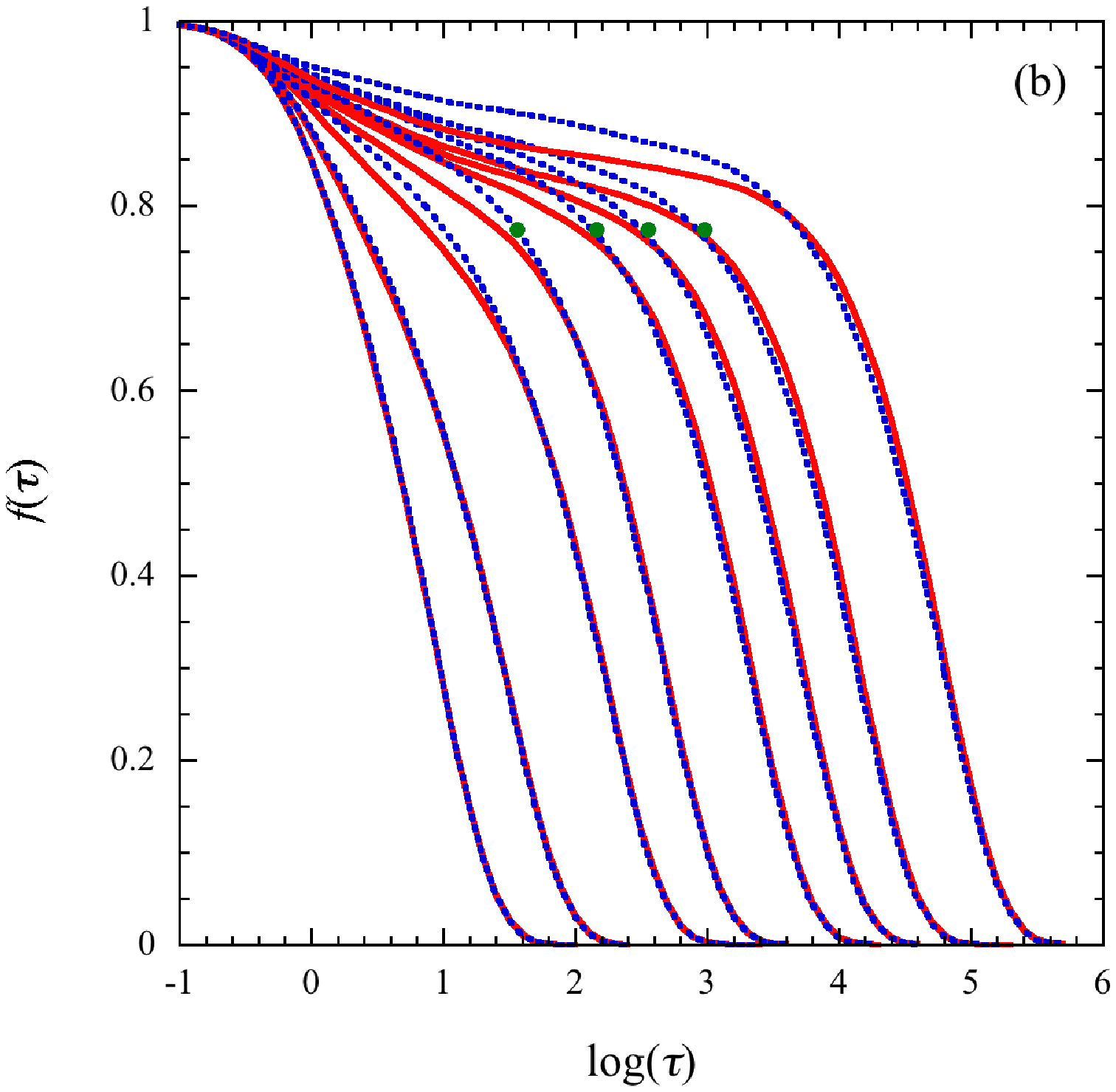}
\end{minipage}
\end{tabular}
\end{center}
\caption{(Color online) A plot of $f(\tau)$ versus $\log(\tau)$ for different temperatures. (a) The dotted lines indicate the numerical solutions for the new model in SW at $T=$[L] 5.00, 1.25, [S$_w$] 0.833, 0.714, [S$_f$] 0.644 ($T_f$), 0.625, 0.556, and [G] 0.527 ($T_g$) from left to right.  The symbols $(\bullet)$ indicate the crossover time in [S$_f$] given by $\log(\tau_x)\simeq 2.002$ ($T=0.644$), 2.306 (0.625), 3.615 (0.556), and 4.931 (0.527), where $f(\tau_x)\simeq 0.682$. (b) The dotted lines indicate the numerical solutions for the new model in NV at $T=$[L] 5500, 4300, [S$_w$] 3600, [S$_f$] 3300, 3000, 2900, [G] 2800, and 2600 (K) from left to right. The symbols $(\bullet)$ indicate the crossover time in [S$_f$] given by $\log(\tau_x)\simeq 1.562$ ($T=3300$K), 2.192 (3000), 2.551 (2900), and 2.978 (2800), where $f(\tau_x)\simeq 0.775$. The solid lines in both figures indicate the numerical solutions for the conventional model.}
\label{comp}
\end{figure}
In Figs. \ref{comp}(a) and \ref{comp}(b), the numerical solutions for the conventional model are now compared directly with those for the new model in SW and NV, respectively. For higher temperatures in [L], both solutions are shown to coincide with each other well within error. As $T$ decreases, the difference between both solutions becomes clear. At $\beta$ stage for $1.0<\tau\leq\tau_x$, the plateau height of $f(\tau)$ for the conventional model is shown to be always lower than that for the new one. At $\alpha$ stage for $\tau_x\leq\tau\leq\tau_{\alpha}$, the distinct difference between both solutions becomes obvious. Here we note that one can also obtain the numerical solutions for the conventional model at any temperatures in [S] since one can find a value of $\varepsilon$ approximately by using a linear equation for $u$. As typical examples, therefore, we also compare the solutions for SW at $T=T_f$ and $T_g$ in Fig. \ref{comp}(a).

\section{Summary}
In this paper, we have proposed the new simplified model given by Eq. (\ref{asmf}) for the nonlinear memory function in the TMCT equation. This model has contained two unknown parameters $\mu$ and $\nu$. In order to fix a value of $\nu$, we have positively used two types of decays, the VS decay and the KWW decay, near the glass transition. Then, we have checked whether the new model can describe the nonlinear memory function obtained numerically by directly solving the ideal TMCT equation given by Eq. (\ref{ka}) with the PY static structure factor or not. Thus, we have shown that the new model can describe it very well, while the conventional model can not for any values of $\varepsilon$. We have first transformed the ideal TMCT equation into an universal form given by Eq. (\ref{kasc}) because there exists only one solution at each value of $u$ for same type of liquids. Hence we have chosen A$_{80}$B$_{20}$ as a typical example of fragile liquids and SiO$_2$ as a typical example of strong liquids. In order to fix a value of $\mu$, we have used the simulation results performed on those systems. We have then solved the universal equation numerically for both models under the same initial conditions. Then, we have compared both numerical solutions together with the simulation results from a unified point of view consistently. Thus, we have shown that the new model can describe the simulation results well not only qualitatively but also quantitatively, except at $\beta$ stage because of the ideal equation, while the conventional model can describe them approximately only at higher temperatures in [L]. 

The new model has suggested that the supercooled state should be further separated into two substates, a weekly supercooled state [S$_w$] and a deeply supercooled state [S$_f$]. In [S$_f$] the nonlinear parameter $\mu$ becomes constant, up to $T_g$. Hence this situation seems to be similar to that in [L] since $\mu$ is constant in [L], up to $T_s$. However, there exists an essential difference between them. In [S$_f$] the magnitude of the equilibrium density fluctuation $\rho(\bm{x},t)$ around $\rho$ is large and $u$ obeys a power law given by Eq. (\ref{sltd}), where $\bm{x}$ is a position vector. On the other hand, in [L] it is small and $u$ is approximately described by a linear equation for $T^{-1}$. In [S$_w$] $\mu$ grows rapidly as $T$ decrease. This situation is similar to that in [G]. However, there also exists an essential difference between them. In [S$_w$] the system is equilibrium and the magnitude of the density fluctuations is larger as $T$ decreases. On the other hand, in [G] the system is out of equilibrium and the magnitude of the nonequilibrium density fluctuation around the average density $\bar{n}(\bm{x},t)$ is small, where $\bar{n}(\bm{x},t)$ obeys a deterministic nonlinear equation and describes a nonequilibrium state. Because of the small fluctuations, therefore, $u$ increases slowly, following the Arrhenius law given by $u\sim T^{-1}$. Those differences may be explored more clearly by using the so-called spatial heterogeneity \cite{ag65,toku97,toku99,si99,ed00,w00,rr02}. As discussed in Refs. \cite{toku97,toku99}, the physical picture is as follows. In [S$_w$] the spatially heterogeneous regions with $|\rho(\bm{x},t)|>\rho$ start to change in space and time at $T$. As $T$ decreases, such regions grow larger, leading to an increment of $\mu$. Among those regions, there are still enough space for particles to diffuse, leading to medium values of a diffusion coefficient $D(T)$. In [S$_f$] the heterogeneous regions seem to almost cover a whole space and change slowly in space and time at $T$. As $T$ decreases, such regions grow slowly, keeping $\mu$ constant. There exist small cages formed by surrounding heterogeneous regions, within which particles can diffuse, leading to smaller values of $D(T)$. In [G] the heterogeneous regions with $|\bar{n}(\bm{x},t)|>\rho$ cover a whole space and change very slowly in space and time at $T$. As $T$ decreases, the heterogeneous regions do not hardly grow, while the nonequilibrium state determined by $\bar{n}(\bm{x},t=0)$ depends on $T$ sensitively, leading to a sharp increment of nonlinearity. There still exist smaller cages for particles to diffuse, following the Arrhenius law. This will be discussed elsewhere.

Finally, we have mentioned that the $u$ dependence of $\mu$ and $\kappa$ for (F) is qualitatively similar to that for (S) but it is quantitatively different from that. This is reasonable because such a difference is originally based on the fact that the static structure factor of SiO$_2$ is structurally quite different from that of fragile liquids because the former has a network structure \cite{z83,m95,kob99}. Hence this confirms that although the universality in the dynamics of glass-forming liquids holds in the same type, the dynamics of different types does not coincides with each other even if $u$ has the same value in each type. In this paper, we have chosen two types of the simulation results as simple examples, SW for (F) and NV for (S). In order to confirm the universal properties and also to obtain precise values of universal quantities, one needs to perform more extensive molecular-dynamics simulations on various glass-forming materials at the same value of $u$ consistently.

Acknowledgments

The author wishes to thank T. Narumi for supplying his unpublished numerical solutions of ideal TMCT equation with the PY static structure factor. This work was partially supported by Institute of Multidisciplinary Research for Advanced Materials, Tohoku University, Japan.

\end{document}